\documentclass[twocolumn,apj]{openjournal}

\usepackage[breaklinks,colorlinks,citecolor=blue,urlcolor=blue]{hyperref}
\usepackage{amsmath}

\usepackage[T1]{fontenc}
\usepackage{newtxtext,newtxmath}
\usepackage{graphicx}
\usepackage{amsfonts}
\usepackage{soul}
\usepackage{enumitem}
\usepackage[hyphens]{xurl}
\usepackage{multirow}

\usepackage{ulem}
\usepackage{natbib}
\usepackage{xcolor}

\definecolor{darkgreen}{rgb}{0.1, 0.6, 0.1}
\definecolor{darkorange}{rgb}{0.8, 0.4, 0.05}

\defcitealias{Oman2021}{O21}
\defcitealias{Karachentsev2013}{K13}

\shorttitle{The abundance of thin dwarf galaxies}
\shortauthors{J. A. Benavides et al.}

\begin{document}
	
\journalinfo{The Open Journal of Astrophysics}
	

\title{\vspace{-0.1cm}The abundance of thin dwarf galaxies: a challenge for cosmological simulations\vspace{-1.5cm}} 
	
\author{
	Jos\'e A. Benavides$^{1}$\href{https://orcid.org/0000-0003-1896-0424}{\includegraphics[scale=0.4]{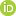}}, %
	Laura V. Sales$^{1}$\href{https://orcid.org/0000-0002-3790-720X}{\includegraphics[scale=0.4]{images/orcid.png}}, 
	Julio F. Navarro$^{2}$\href{https://orcid.org/0000-0003-3862-5076}{\includegraphics[scale=0.4]{images/orcid.png}},
	Simon D. M. White$^{3}$\href{https://orcid.org/0000-0002-1061-6154}{\includegraphics[scale=0.4]{images/orcid.png}},
	Carlos S. Frenk$^{4,6}$\href{https://orcid.org/0000-0002-5908-737X}{\includegraphics[scale=0.4]{images/orcid.png}},\\
	Kyle A. Oman$^{4,5,6}$\href{https://orcid.org/0000-0001-9857-7788}{\includegraphics[scale=0.4]{images/orcid.png}}
	and Shaun Cole$^{4,6}$\href{https://orcid.org/0000-0002-5954-7903}{\includegraphics[scale=0.4]{images/orcid.png}}
	\\
	\vspace{0.05cm}
	$^{1}$Department of Physics and Astronomy, University of California, Riverside, CA, 92507, USA\\ 
	$^{2}$Department of Physics and Astronomy, University of Victoria, Victoria, BC V8P 5C2, Canada\\ 
	$^{3}$Max-Planck-Institut für Astrophysik, Karl-Schwarzschild-Straße 1, 85748 Garching, Germany\\
	$^{4}$ Institute for Computational Cosmology, Durham University, South Road, Durham DH1 3LE, United Kingdom\\
	$^{5}$ Centre for Extragalactic Astronomy, Durham University, South Road, Durham DH1 3LE, United Kingdom\\
	$^{6}$ Department of Physics, Durham University, South Road, Durham DH1 3LE, United Kingdom
}

\thanks{$^\star$E-mail:} \email{joabenavid@ucr.edu}
\email{jose.astroph@gmail.com}

\begin{abstract} 
    We study the prevalence of thin galaxies as a function of stellar mass in the range $10^7<M_{\star}/\rm{M_\odot}<10^{11}$ using data from the GAMA, DESI, ALFALFA and Nearby Galaxy catalogs. We use the distribution of projected axis ratios, $q$, to infer the abundance of intrinsically flat galaxies needed to reproduce the observed abundance of highly elongated systems in projection. We find that as many as $40\%$ of galaxies in the mass range $10^9<M_{\star}/\rm{M_\odot}<10^{10}$ are intrinsically flatter than $1$:$5$ (i.e., $c/a<0.2$), a fraction that rises to $\sim 80\%$ for $c/a<0.3$. Although the incidence of thin galaxies decreases towards lower and higher $M_{\star}$, they are still quite common in dwarfs: $\sim 30\%$ and $\sim 65\%$ of $\sim 10^8 ~ \rm{M_\odot}$ galaxies are inferred to be intrinsically flatter than $c/a=0.2$ and $0.3$, respectively. A comparison of these results with several state-of-the-art cosmological hydrodynamical simulations (TNG50, FIREbox, Romulus25)  reveals a distinctive lack of thin simulated dwarfs. In particular, there are no $M_{\star} < 10^9 ~ \rm{M_{\odot}}$ simulated galaxies flatter than $c/a=0.2$, in clear contrast with observational samples. This discrepancy likely reflects limitations in resolution and in the treatment of baryonic physics, suggesting that our understanding of the mechanisms regulating the formation of disk galaxies less massive than the Milky Way is still quite incomplete. Our results present a clear challenge to current numerical models of dwarf galaxy formation, which future models should attempt to meet.

\end{abstract}

\maketitle


\section{Introduction}
\label{SecIntro}

The formation of disk galaxies, like the Milky Way (MW), has long been considered a challenge in the current $\Lambda$CDM paradigm of hierarchical structure formation. And with good reason: disk galaxies are common among systems with stellar mass comparable to the MW, but for a long time their masses, morphologies, and rotation velocities were quite difficult to reproduce in cosmological hydrodynamical simulations \citep[e.g.,][]{NavarroBenz1991,SteinmetzNavarro2002,Samland2003, Scannapieco2009}. The solution to this challenge was enabled by a deeper understanding of stellar feedback and, more specifically, of how it couples to the surrounding gas to delay star formation and prevent the accumulation of low-angular momentum gas by generating powerful galactic outflows. As a result, many state-of-the-art simulations are now almost routinely able to form galaxies that look remarkably like the MW \citep[e.g., ][]{Governato2008,Brook2011,Guedes2011,Agertz2011,Marinacci2014,Hopkins2014,Grand2017,Ferrero2021}.

It is also well established that galaxy disks become less common at larger stellar masses ($M_{\star} \geq 10^{11} ~ \rm{M_{\odot}}$). This can be readily accommodated in $\Lambda$CDM, where the hierarchical assembly of gas-poor galaxies promotes the formation of early-type dispersion-dominated systems in massive halos \citep[e.g., ][]{Fall1979,Steinmetz2002,Meza2003,Sales2012,Naab2014}. On the other hand, the incidence of disks at the other end of the mass spectrum, in dwarf galaxies much fainter than the MW, is much less well established. Since Edwin Hubble's pioneering efforts to classify galaxies \citep{Hubble1926}, low-mass galaxies have been associated with irregular morphologies and therefore decoupled from the main ``tuning fork'' of the Hubble sequence.

In addition, the faintest galactic systems known (the faint and ultra-faint satellites of the Milky Way and Andromeda galaxies) are all dwarf spheroidal (dSph) galaxies with little sign of coherent rotation. Does this mean that thin stellar disks cannot form below a certain stellar mass? If disks are common on the scale of the MW, but vanish in the regime of dwarfs, at what mass scale does this morphological transition occur and how gradual is it?  A robust characterization of the transition should provide valuable constraints to current dwarf galaxy formation models.

Observational limitations make these questions challenging to address. Dwarf galaxies are faint, and complete unbiased samples of dwarf galaxies were difficult to assemble over the limited volumes covered by early surveys. For example, in the Local Group and nearby galaxies, our best laboratory for dwarf galaxy studies, only 31 out of 96 galaxies with $M_{\star}< 10^{9} ~ \rm{M_{\odot}}$ are dwarf irregulars (dIrrs), and $\sim 68\%$ of them are spheroidals (dSphs) that lack disks \citep{McConnachie2012,Roychowdhury2013}. A similar result has been reported in galaxy clusters, such as Virgo \citep{Binggeli1995} and Fornax \citep{Davies1988,Ferguson1989}, where ``spheroidal'' and ``elliptical'' systems (i.e., not disk-dominated) appear to be the dominant populations. These samples, however, have been collected over small volumes in relatively high-density environments, so it is unclear how representative these fractions are when averaged over a cosmologically significant volume that includes isolated, field dwarfs.

An effective way to quantify morphology in large galaxy samples is through the study of galaxy shapes.  Measuring the projected axis ratios $q=b_{\rm p}/a_{\rm p}$ of the luminous component (where $a_{\rm p}$ and $b_{\rm p}$ are the semimajor and semiminor axes of the best-fitting elliptical photometric model), and assuming random projections on the sky, the intrinsic (3D) axis ratio distribution with $a > b > c$ can be inferred, under the assumption that most galaxies are either oblate or prolate \citep[e.g., ][]{Lambas1992,Padilla2008,Chang2013,Rodriguez2016,KadoFong2020}. If galaxies are oblate (i.e., $a=b>c$),  low $q$ values correspond to nearly edge-on disks, whereas $q \sim 1$ can reflect both truly spheroidal objects as well as thin disks seen nearly face-on.

Complicating factors include the fact that measurements of shape often vary with wavelength, and are sensitive to the presence of dust and to the exact radius at which shapes are measured. Most of these effects tend in general to bias the projected shape distributions toward larger $q$ values, implying that the abundance of highly ellipsoidal (low $q$) objects can provide reliable lower limits on the prevalence of highly flattened systems. Strictly speaking, studies of galaxy shapes constrain only the 3D flattening of such systems, but, as we discuss  in Sec.~\ref{SecDiscussion}, highly  flattened systems are most likely thin, rotation-dominated disks.  Because of this, we shall use ``thin disks'' and ``flat galaxies'' almost interchangeably throughout this manuscript.

Previous studies of galaxy shapes have argued that galaxies become more spheroidal as we move toward fainter galaxies \citep[e.g., ][and references therein]{SanchezJanssen2010}. A similar trend is also apparent when considering the shape of gaseous, rather than stellar, disks \citep{StaveleySmith1992,Roychowdhury2010}. These findings appear in qualitatively good agreement with cosmological galaxy formation simulations, which have also reported a well-defined mass-morphology relation where galaxies become increasingly more dispersion-dominated toward lower masses \citep{ElBadry2018a,Zeng2024,Celiz2025,Benavides2025b}. However, the scale, shape, and origin of the transition seems to vary from model to model, suggesting that the disk morphologies of low-mass galaxies are still not fully understood.

From a theoretical perspective, there are several factors that can contribute to making dwarf galaxies thicker than their more massive counterparts. For the gas component, the coupling of feedback to the surrounding interstellar medium can drive ``turbulent'' motions and cycles of gas outflow/inflow  \citep{ElBadry2016}. The periodic removal of gas from the central regions due to outflows can also drive gravitational potential fluctuations that heat-up and expand collisionless components, like dark matter and stars \citep{Navarro1996b,Pontzen2012,Mercado2021,Riggs2024}. Even substructure in the dark matter halo may play a role, where disks could potentially be heated up by the tidal effects and collisions of massive subhalos \citep{Helmi2012}. 

Yet we also know that there is a sizable population of dwarf galaxies (i.e., $M_{\star} < 10^{9} ~~ \rm{M_\odot}$) that exhibits remarkably high rotation, which suggests that the impact of feedback and substructure might not be as deleterious to the formation of rotationally supported disks in dwarf galaxies as reported by the aforementioned simulations. For instance, the majority of the {\sc LITTLE THINGS} sample shows clearly rotation-dominated gas disks with $V_{\rm rot}/\sigma \sim 4 \rm - 10$ and low vertical velocity dispersion $\sigma_z \sim 10 ~ \rm{km \, s^{-1}}$ \citep{Oh2015,Iorio2017}. The dwarfs in the sample of \citet{ManceraPina2025} provide further examples of this population. 

The presence of rotationally-supported gas disks in present-day dwarfs does not automatically imply that stellar thin disks in dwarfs should exist, but they suggest that such disks may very well be present in dwarfs. There is also anecdotal evidence for very thin edge-on disky dwarfs in the literature. Examples include dwarfs like UGC 7321, with an estimated stellar mass of $3 \times 10^8 ~ \rm{M_\odot}$ and axis ratio $q \sim 0.1$ \citep{Banerjee2010} as well as a number of faint dwarfs ($B$-band magnitudes $M_B>-19$) in the Catalogue of Flat Galaxies \citep{Karachentsev1989}. Even closer to home, the Triangulum galaxy M33, with $M_{\star} \sim 5 \times 10^9 ~ \rm{M_\odot}$ is another example of a cold stellar disk with $V_{\rm rot}/\sigma > 6$ \citep{Quirk2022}.

In this work, we use a large sample of observed galaxies to revisit the prevalence of thin disks in dwarf galaxies. We include galaxies from the GAMA and DESI catalogs \citep{Driver2009, DESICollaboration2025}, which provide the largest statistical base for the analysis. We also use results from the Catalog of Nearby Galaxies \citep{Karachentsev2013}, which help us explore how biased the local galaxy samples are relative to a more representative volume. Lastly, we also consider a sample based on the cross match of the ALFALFA catalog \citep{Haynes2018,Durbala2020} with the Sloan Digital Sky Survey \citep[SDSS,][]{Abazajian2009}. This addition is important as it is an HI-selected sample and therefore free of the many biases incurred by optical selection in the other catalogs. Our main results regarding the incidence of flat dwarf galaxies are then compared with a few recent cosmological hydrodynamical simulations. Such simulations are typically tuned to fit the galaxy mass function, as well as the size, and general morphology of massive galaxies, but have not been calibrated  to the morphologies of dwarfs, making this comparison an instructive test of the adequacy of current dwarf galaxy formation simulations.

This paper is organized as follows: in Sec.~\ref{SecData} we describe the observational catalogs and simulations. In Sec.~\ref{SecSample} we take a closer look at the observed galaxy sample and study their projected shape distributions. Sec.~\ref{SecModel} introduces the deprojection method used to estimate the intrinsic distribution of axis ratios $c/a$ that we later apply to both observations and simulations. Our main results regarding intrinsic galaxy shapes are discussed in Sec.~\ref{SecDiscussion}. We summarize our findings in Sec.~\ref{SecConc}. 

\section{Galaxy Samples}
\label{SecData}

\subsection{Observations}
\label{SSecObs}

We use in this paper mainly data for galaxies in the range of stellar mass $10^7<M_{\star}/ \rm{M_\odot}<10^{11}$  and redshift $z<0.1$, collected from different observational surveys. Details of each survey are presented briefly below.

\subsubsection{GAMA}
\label{SSSecGAMA}

The Galaxy And Mass Assembly (GAMA\footnote{\href{https://www.gama-survey.org/}{https://www.gama-survey.org/}}) Survey \citep{Driver2009} survey was designed to study the formation and evolution of galaxies in the nearby Universe ($\sim300,000$ galaxies in the redshift range $0\lesssim z \lesssim 0.25$), with data taken at the Anglo-Australian Telescope (AAT), using the AAOmega spectrograph. We use data from GAMA's DR3 \citep{Baldry2018}; more specifically the data listed in the StellarMassesv20 \citep{Taylor2011} and SersicCatSDSSv09 \citep{Kelvin2014} catalogs. The imaging data from the GAMA survey reach surface brightness limits down to $\sim 29 ~ \rm{mag \, arcsec^{-2}}$ in the r-band \citep{Driver2011, Driver2016, Williams2016}.

GAMA galaxy shapes are computed from 4-parameter S\'ersic fits (SersicCatSDSS.fits) to the surface brightness profiles of galaxy images in the $r$-band. The photometric fits yield a mean ellipticity, $\epsilon=1-q$, a characteristic surface brightness and radius, as well as a S\'ersic index $n_{S}$. We use here the values of $q$ derived from the best fit (as listed in column ``GALELLIP\_r'' of Table SersicCatSDSS of catalog SersicPhotometry/v09), as well as the total stellar mass of the galaxy, $M_{\star}$, estimated from the galaxy apparent luminosity, colors, and redshift. Stellar masses are listed in column ``logmstar'' of Table StellarMasses of catalog StellarMasses/v20.

We have used the GAMA sample to explore the dependence of the measured shapes on photometric band, including the $u$, $g$, $r$ and $i$ bands. Overall, the shapes measured in all bands show a strong 1:1 correlation with the shapes measured in the $r$ filter used in our analysis. The rms dispersion from such correlations are 0.143, 0.051 and 0.052 for the $u$, $g$ and $i$ bands, respectively. Shapes in the $u$ band are more strongly influenced by young stellar populations, ongoing star formation, and atmospheric absorption, explaining the larger dispersion and poorer agreement with the other bands.

\subsubsection{DESI}
\label{SSSecDESI}
The Dark Energy Spectroscopic Instrument (DESI) survey is a 5-year spectroscopic redshift survey that has produced a detailed map of the evolving structure of the Universe in the redshift range $0 \leq z \lesssim 4$ \citep{DESICollaboration2025}. We use the open access DESI\footnote{\href{https://www.desi.lbl.gov/}{https://www.desi.lbl.gov/}} DR1\footnote{\href{https://data.desi.lbl.gov/public/dr1/vac/dr1/stellar-mass-emline/}{https://data.desi.lbl.gov/public/dr1/vac/dr1/stellar-mass-emline/}}, which consists of data collected during the first year of the DESI main survey. The dataset includes high-confidence redshifts for $\sim13.7$ million objects classified as galaxies. Surface brightness limits in DESI reach roughlyn $\sim 28.5$-$29 ~ \rm{mag \, arcsec^{-2}}$ \citep{MartinezDelgado2023, Kwon2025} in the r-band.

DESI galaxy shapes ($q$) are computed from the ellipticity through the complex numbers $\epsilon= \epsilon_1 + i \epsilon_2$, explicitly:

\begin{equation}
q = \frac{1 - |\epsilon|}{1 + |\epsilon|} \, ,
\label{eq:e_desi}	
\end{equation}

\noindent where $|\epsilon| = \sqrt{ \epsilon^2_1 + \epsilon^2_2}$. The ellipticity components ($\epsilon_1$, $\epsilon_2$) are listed in the columns ``SHAPE\_E1'' and ``SHAPE\_E2'' of Table dr1\_galaxy\_stellarmass\_lineinfo\_v1.0 of the catalog stellar-mass-emline/v1.0. Stellar masses are derived using Code Investigating GALaxy Emission \citep[CIGALE;][]{Boquien2019, Yang2020}, which combines broad-band photometry from the DESI Legacy Imaging Surveys (g, r, z, W1, and W2), with spectrophotometric measurements obtained from ten artificial bands generated by convolving the DESI spectra with the corresponding filter responses. These stellar masses are listed in column ``MASS\_CG'' of this catalog.

\vspace{0.4cm}

\subsubsection{ALFALFA}
\label{SSSecALFLFA}
The Arecibo Legacy Fast ALFA (ALFALFA\footnote{\href{https://egg.astro.cornell.edu/}{https://egg.astro.cornell.edu/}}) Survey \citep{Giovanelli2005}, provides HI 21-cm line measurements for $\sim 31,500$ galaxies over nearly $7000 ~ \rm{deg^2}$ on the sky. More specifically, we use the catalog presented in \citet{Haynes2018} and \citet{Durbala2020}, complemented with data from the SDSS DR7 \citep{Abazajian2009}. Sources were cross-identified in the two catalogs using their positions in the sky \citep[see ][]{Oman2021}. The surface brightness limits in this catalog are set by SDSS imaging, which typically reaches $\sim 28.5 ~ \rm{mag \, arcsec^{-2}}$ in the r-band \citep{Fliri2016}, or $\sim 27.9 ~ \rm{mag \, arcsec^{-2}}$ in g \citep{Janowiecki2015}. 

The main reason for including the ALFALFA catalog in our analysis is its very different selection function compared to the other surveys included here: objects in ALFALFA are blind to stellar properties and are selected, instead, by a lower limit in 21cm flux. Optically selected samples tend to under-represent face-on galaxies, since a less inclined disk has a lower observed surface brightness (ignoring dust effects). In contrast, HI-selected samples are biased against edge-on systems because, at fixed total flux, the emission is distributed across more channels for an edge-on system, reducing the S/N.

The shape of galaxies in ALFALFA is obtained from the $q$ (in the $r$-band) of SDSS, using the axial ratio of the best fitting 2D elliptical exponential model. The $q$ values are listed in column ``expAB\_r'' of Table 1 of the \citet{Durbala2020} catalog\footnote{Availabe in \href{https://egg.astro.cornell.edu/alfalfa/data/index.php}{https://egg.astro.cornell.edu/alfalfa/data/index.php}}. The stellar mass is taken from SDSS optical photometry, in logarithmic solar units using the methods of \citet{Taylor2011}, as listed in column ``logMstarTaylor'' of Table 2 of \citet{Durbala2020} catalog. We only use 28,267 galaxies (of 31,501) marked with sdssPhotFlag=1 \citep[equivalent to the `Code 1' of][]{Haynes2018}, which have uncertainties of less than 0.05 mag in g and i \citep[see Sec. 6 of][]{Durbala2020}. This ALFALFA catalog is limited to $z \lesssim 0.06$.

\subsubsection{Nearby Galaxy Catalog}
\label{SSSecK2013}

We also include information from the all-sky catalog of 869 nearby galaxies from the Catalog of Neighboring Galaxies from \citet{Karachentsev2013}. These are galaxies within 11 Mpc and with recession velocity  $V_{\rm LG} \leq 600 ~ \rm{km \, s^{-1}}$ relative to the Local Group barycenter. This catalog includes many Local Group satellites and other very low-mass objects, which are historically important due to their proximity and to the high quality data available to study them. 

The shape of the galaxies in this catalog is obtained directly from Table~1 of \citet{Karachentsev2013}, which provides information on the apparent axial ratio measured at the Holmberg isophote ($\sim 26.5 ~ \rm{mag ~ arcsec^{-2}}$) in the $B$-band. The stellar mass of these galaxies is calculated from the luminosity, in the $K_s$-band, presented in Table 2. This luminosity has been corrected for extinction \citep[for details see Sec. 3 of][]{Karachentsev2013}. We compute stellar masses assuming a mass-to-light ratio of $\rm{M_\odot L_\odot^{-1}} = 0.6$ in the $K_s$-band following \citet{McGaughSchombert2014}.

\begin{figure}
	\centering
	\includegraphics[width=\columnwidth]{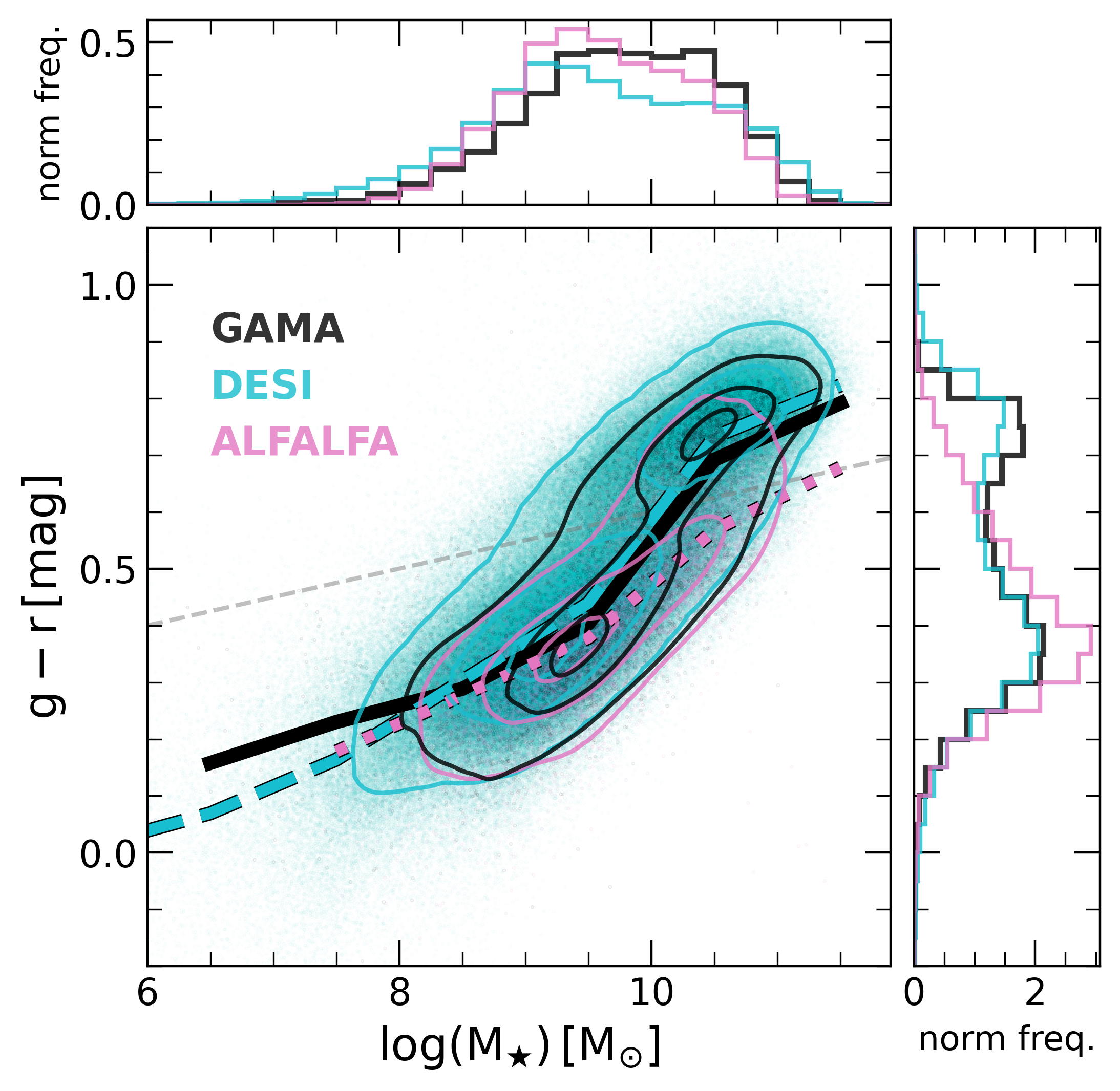}
	\caption{Color ($g-r$) as a function of stellar mass for all galaxies included in our GAMA (black), DESI (cyan) and ALFALFA (pink) samples. The main panel shows individual galaxies as dots along with $10\%$, $50\%$ and $90\%$ percentile contours, as well as thick curves tracing the median color as a function of $M_{\star}$. Horizontal and vertical histograms summarize the distributions in stellar mass and colors for each sample. A gray dashed line roughly traces the $M_{\star}$-dependent color relation that splits the region between the red and blue peaks (i.e., the color bimodality) in the main panel. Overall, our sample includes a fair representation of red and blue galaxies, see text for more details.}
	\label{fig:color_mstar}
\end{figure}

\begin{figure}
	\centering
	\includegraphics[width=\columnwidth]{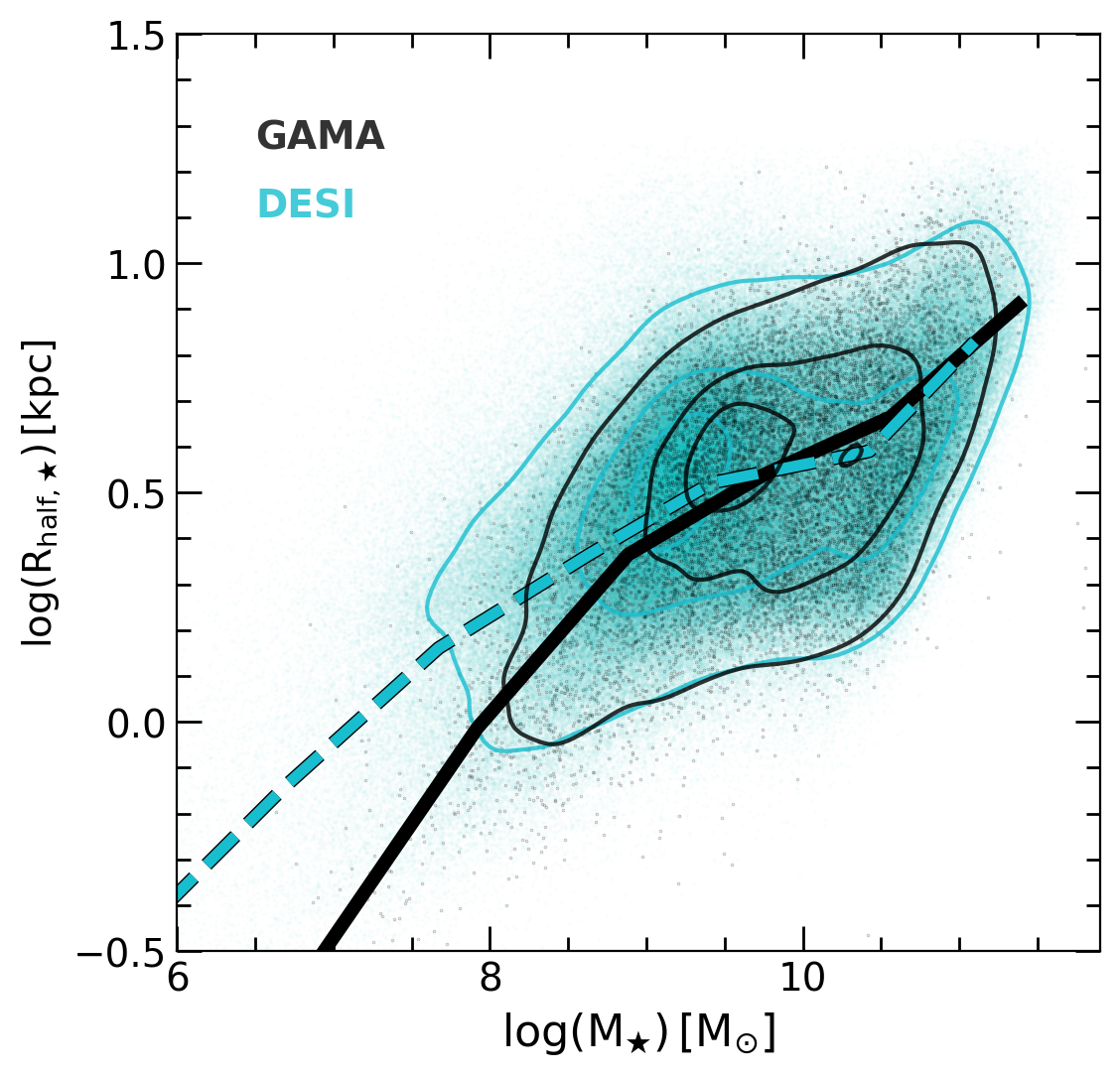}
	\caption{Stellar half-light radius ($R_{\rm{half,\star}}$) as a function of stellar mass for our GAMA  (black) and DESI (cyan) samples. There are no large differences in the size of objects included in our GAMA and DESI sample with $M_{\star} \gtrsim 10^9 ~ \rm{M_{\odot}}$, although DESI shows some preference towards larger dwarfs for low-mass objects, $M_{\star} < 10^9 ~ \rm{M_{\odot}}$.} 
	\label{fig:mstar_reff}
\end{figure}

\subsection{Simulations}
\label{SSecSims}

We compare our main results with  data for samples of simulated galaxies taken from  the literature. We include data from three different recent cosmological hydrodynamical simulation projects: TNG50, FIREbox, and Romulus25, briefly described below.

\subsubsection{TNG50}
\label{SSecTNG50}

The TNG50-1 (hereafter TNG50 for short) cosmological hydrodynamical simulation \citep{Pillepich2018a, Pillepich2018b, Nelson2018, Pillepich2019, Nelson2019TNG} evolves a cosmological volume approximately 50 Mpc on a side using the {\sc arepo} code \citep{Springel2010}. This simulation adopts cosmological parameters  consistent with the measurements from \citet{PlankColaboration2016}: $\Omega_{\Lambda} = 0.6911$, total matter content (dark matter + baryons) $\Omega_{\rm m} = \Omega_{\rm drk} + \Omega_{\rm bar} = 0.3089 $, $\Omega_{\rm bar} = 0.0486$, Hubble constant $H_0 = 100 \, h \, $km$ \, $s$^{-1} \, $Mpc$^{-1}$, $h = 0.6774$, $\sigma_8 = 0.8159$, and spectral index $n_s = 0.9667$.

The dark matter particle mass resolution is ${m_{\rm drk}} = 4.5 \times 10^5 ~ \rm{M_{\odot}}$, while the typical baryonic element mass (gas cells or star particles) is $m_{\rm bar} \sim 8.5 \times 10^4 ~ \rm{M_{\odot}}$. The gravitational softening for both dark matter and stars is set to $\epsilon = 0.29$ kpc at $z=0$, with the softening becoming significantly smaller for gas cells in high-density regions, reaching $50\, h^{-1}$ pc in the highest-resolution regions. Our sample includes all central (i.e., no satellites) galaxies in the stellar mass range of interest, $10^7<M_{\star}/\rm{M_{\odot}}< 10^{11}$. To compute the shape of simulated galaxies in TNG50, we generate a random projection of the stellar particles associated to each galaxy of interest. We then calculate the projected axis lengths ($a_p$, $b_p$) through the eigenvalues of the 2D stellar mass-weighted inertia tensor, including all stellar particles within $3$ times the half-mass radius of the stars (i.e., $r<3 r_{\rm half, \star}$).

\subsubsection{FIREbox}
\label{SSecFIREbox}

We use the FIREbox \citep{Feldmann2023} cosmological simulation, which is part of the Feedback In Realistic Environments (FIRE\footnote{\href{https://fire.northwestern.edu/}{https://fire.northwestern.edu/}}) project \citep{Hopkins2014, Hopkins2018}. FIREbox follows the evolution of a ($22.1\,$Mpc$^3$) volume set up initially with $2 \times 1024^3$ gas elements and dark matter particles.  The mass resolution of baryons is $m_{\rm bar} = 6.3 \times 10^4 ~ \rm{M_\odot}$ and $m_{\rm DM} = 3.3 \times 10^5 ~ \rm{M_\odot}$ for dark matter particles, with a gravitational softening of $12$ pc for stars and $80$ pc for the dark matter. The force softening of gas cells is adaptive, reaching a minimum of 1.5 pc in the dense ISM. Unlike TNG50, FIREbox follows the evolution of multi-phase gas, modeled using the FIRE-2 framework. We highlight the small softening length in the FIREbox simulation compared to the other simulations used in this work and briefly discuss its impact in App.~\ref{app:resolution}.

The initial conditions are generated at redshift $z = 120$ using the MUlti-Scale Initial Conditions \citep[MUSIC,][]{music2011} code. It assumes a set of cosmological parameters consistent with the \citet{PlankColaboration2016} measurements: $\Omega_{\rm m} = \Omega_{\rm dm} + \Omega_{\rm bar} = 0.3089, \Omega_{\rm bar} = 0.0486$, $ \Omega_{\rm \Lambda} = 0.6911$, Hubble constant $\rm{ H_0 = 100 \, h \, km \, s^{-1} \, Mpc^{-1} }$, $ h = 0.6774 $, $\sigma_8 = 0.8159 $ and spectral index $ n_s = 0.9667$.

We use in this study the distribution of galaxy shapes presented in \citet[more specifically, their Figs. 4 \& 5]{Klein2025}. This data corresponds to a sample of central galaxies in the range of stellar mass $10^8<M_{\star}/\rm{M_\odot}<10^{11}$. Shapes were calculated using SDSS $r$-band mock images of FIREbox galaxies, a procedure quite similar to what is adopted in the observational surveys mentioned above. 

\subsubsection{Romulus25}
\label{SSecRomulus}

We also use here measures of galaxy shapes from the Romulus25 simulation \citep{Tremmel2017}, which evolves a cosmological box of 25 Mpc on a side. This cosmological hydrodynamical simulation uses the N-body + smoothed particle hydrodynamics (SPH) code CHANGA \citep{Menon2015}. Romulus25 adopts cosmological parameters consistent with the \citet{Planck2014} with $\Lambda$CDM cosmology $\Omega_{\rm m} = \Omega_{\rm dm} + \Omega_{\rm bar} = 0.3086$, $\Omega_{\rm \Lambda} = 0.6914, h = 0.6777 $, and $\sigma_8 = 0.8288$. The mass resolution of baryons is $m_{\rm bar} = 2.12 \times 10^5 ~ \rm{M_\odot}$ and $m_{\rm DM} = 3.39 \times 10^5 ~ \rm{M_\odot}$ for dark matter particles. Gravitational interactions are resolved with a softening length of $350$ pc.

The 3D shapes of ROMULUS galaxies are determined from the eigenvalues of the shape tensor, using the stellar particles located within twice the half-light radius. From these shapes, a mock 3D ellipsoid is generated based on the $b/a$ and $c/a$ axis ratios for 100 random orientations. This ellipsoid is then projected onto a 2D plane to measure the projected ellipticity.  We use the projected shape ($q$) distribution of simulated galaxies as reported in  Fig. 8 of \citet{VanNest2022}, considering only  isolated galaxies, and converting the listed ellipticities ($\epsilon$) into axis ratios using $q = 1 - \epsilon$.\\

\begin{figure}
	\centering
	\includegraphics[width=\columnwidth]{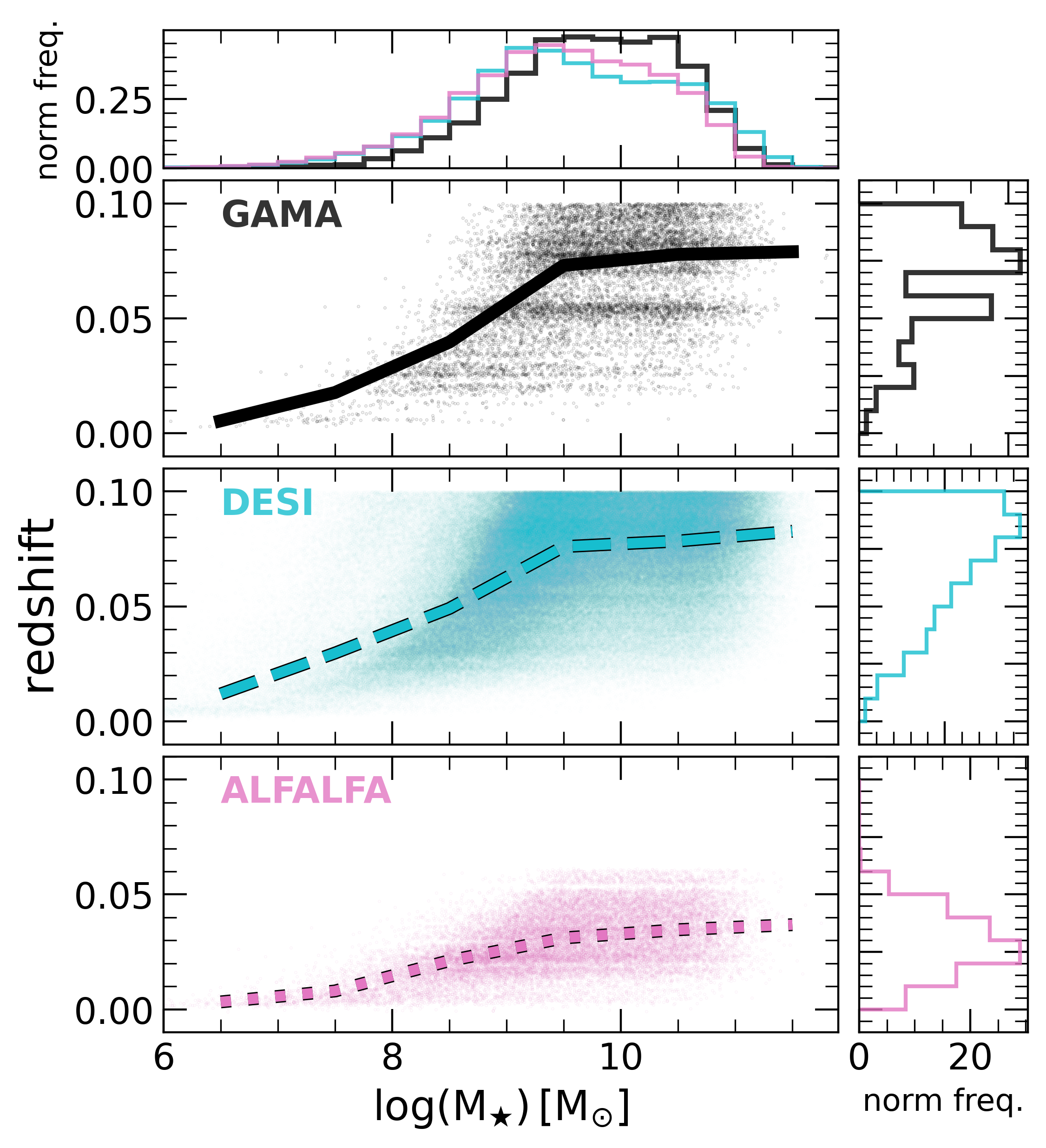}
	\caption{Stellar mass vs. redshift of galaxies in the observational sample. Dots are used for individual galaxies while thick lines show the median trend as a function of $M_{\star}$. The ALFALFA catalog (pink) samples a more local volume than that covered by GAMA (black) and DESI (cyan). The redshift range spanned by  our samples depend on $M_{\star}$, although the only explicit cut used to select them was $z<0.1$. Histograms along the axes help to visualize the stellar mass (horizontal) and redshift (vertical) distributions.}
	\label{fig:mstar_z}
\end{figure}

\begin{figure*}
	\centering
	\includegraphics[width=1.0\textwidth]{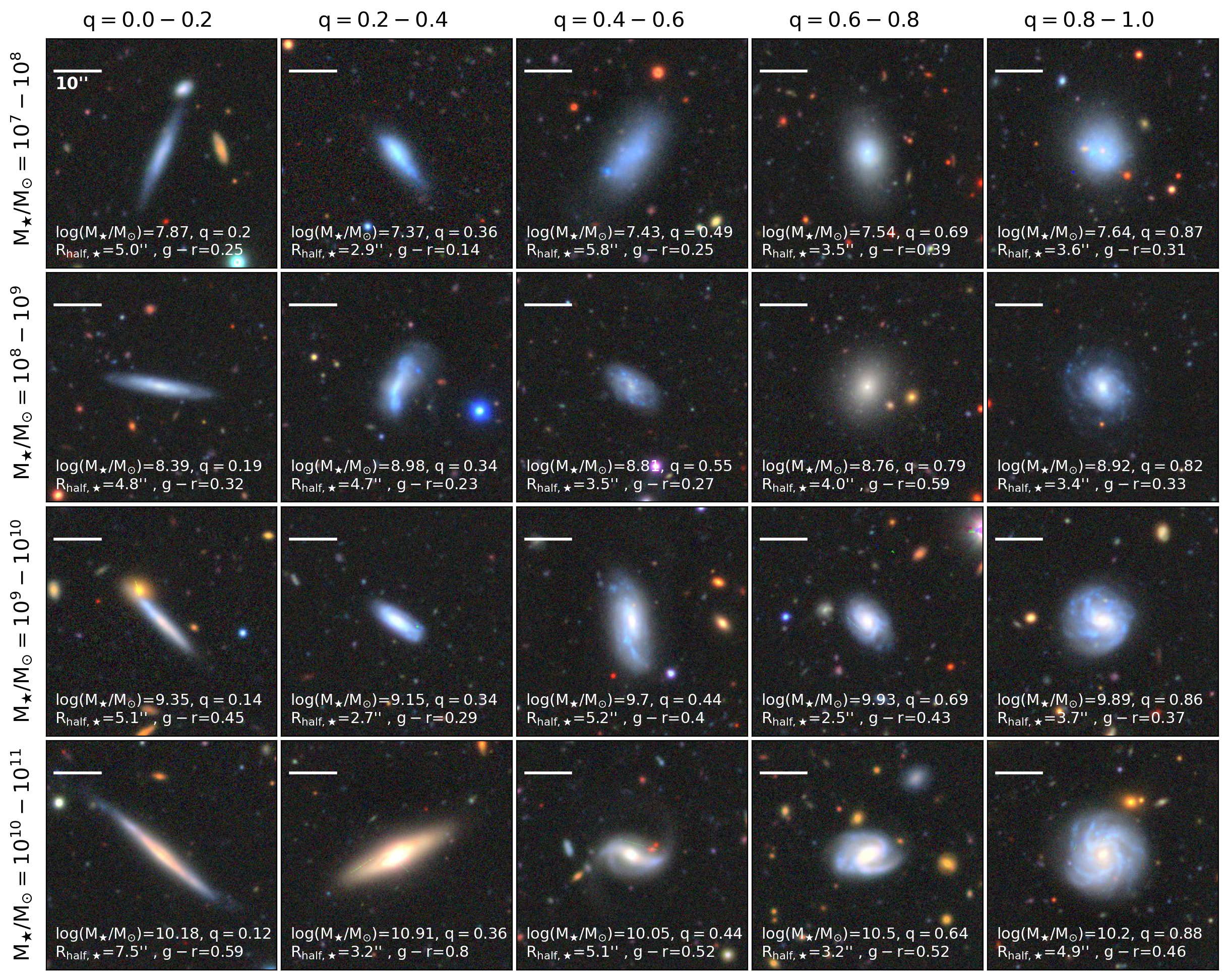}
	\caption{Examples of galaxies with different projected axis ratio $q$ (columns) in bins of stellar masses (rows) from our DESI sample. Low $q$ values are quite clearly associated with edge-on disks, while larger $q$ can be associated to thicker/rounder objects and also face-on disks. Note the presence of well defined thin edge-on disks even in our lowest mass systems (left column, top two rows). All images were generated using the Legacy Survey database.}
	\label{fig:stamps}
\end{figure*}

\section{Observed Galaxy Shapes}
\label{SecSample}

\subsection{Main characteristics of the galaxy sample}

Our galaxy sample includes all galaxies in the mass range $10^7<M_{\star}/\rm{M_\odot}< 10^{11}$ listed in the catalogs described in the previous section, with the additional restriction that we consider only galaxies with redshift $z<0.1$. Fig.~\ref{fig:color_mstar} shows the stellar mass vs ($g-r$) color relation for all galaxies in our GAMA (black), DESI (cyan) and ALFALFA (pink) samples.

For GAMA galaxies we estimate their colors from the ratio between the absolute luminosities in the $g$ and $r$ rest frame bands (as listed in the StellarMassesv20 catalog). For DESI galaxies we estimate the colors using the legacy survey fluxes (FLUX\_G, FLUX\_R). Since the two catalogs have slightly different calibrations we have corrected the DESI colors by a small amount to make them consistent with GAMA colors. The correction is color-dependent, and is derived by comparing the colors of  all galaxies in common in both GAMA and DESI. The correction has the form: 
\begin{equation}
f(X_{g-r}) = \alpha_{g-r} X_{g-r} + \beta_{g-r} \, ,
\label{eq:color_corr}	
\end{equation}

\noindent where, $\alpha_{g-r} = 1.073$ and $\beta_{g-r} = 0.082$ for GAMA to DESI color correction. In the case of ALFALFA we include in Fig.~\ref{fig:color_mstar} the $M_{\star}$ vs $(g-r)$ relation for galaxies in the \citet{Oman2021} catalog, without further correction. For each catalog, we show individual galaxies with small dots, contours to characterize the distribution of galaxies in the stellar mass-color plane, as well as thick lines to trace the median colors as a function of $M_{\star}$. The side panels show the color and $M_{\star}$ distributions of each galaxy sample.

The DESI and GAMA galaxy samples show the familiar red/blue color bimodality \citep[see, e.g.,][]{Baldry2004, Baldry2006, Faber2007, Ball2008, OMill2008}, which is also apparent from the color histograms in the right side panel.  The two sequences separate cleanly when split according to an $M_{\star}$-dependent color, traced by the gray dashed line in the figure. The color bimodality becomes weaker for $M_{\star} < 10^{9.5} ~ \rm{M_\odot}$, with most fainter dwarfs having only relatively blue colors, albeit with a large scatter. As expected, ALFALFA galaxies undersample the red sequence because of the HI selection. As a result,  the ALFALFA sample is bluer than GAMA and DESI at the massive end.

Fig.~\ref{fig:mstar_reff} compares the stellar mass vs size relation for the GAMA (black) and DESI (cyan) galaxies in our sample. Sizes are estimated as follows. For GAMA, we use the effective half-light radius ($R_{\rm{half},\star}$) in the $r$-band (GALRE\_r) estimated by using GALFIT to fit the $r$-band photometry. For DESI, we use the half-light radius (SHAPE\_R) from the Legacy Survey photometric catalog.

Since we are interested in galaxy shapes, we impose a minimum angular size ($R_{\rm{half},\star}>1$ arcsec) to minimize the effects of seeing on the measurements. We also impose a maximum angular size ($R_{\rm{half},\star}<10$ arcsec) in order to remove galaxies that are artificially fragmented, or affected by other artifacts. These restrictions remove a relatively small number of galaxies from our sample, 81,110 (or $11.7\%$), out of a total of 690,631 for DESI, and 1,290 (or $10.3\%$) out of a total of 11,498 for GAMA.

We express the stellar half-light radius in kiloparsecs using their angular size and redshift, assuming the \texttt{Planck15} cosmology \citep{PlankColaboration2016} as implemented in  Astropy\footnote{\href{https://docs.astropy.org/}{https://docs.astropy.org/}} \citep{Astropy2018}.  
The two galaxy sets compare well at the massive end, but there is some indication that GAMA galaxies are smaller at the low-mass end (i.e., $M_\star \lesssim 10^9 ~ \rm{M_\odot}$). This is probably due to the higher sensitivity of DESI at low surface brightness. 

Although we apply a uniform cut in redshift ($z<0.1$) to all of our samples, this results in different effective redshifts for different stellar masses. We show in Fig.~\ref{fig:mstar_z} the $M_{\star}$ vs $z$ relation for GAMA, DESI and ALFALFA. Individual galaxies are shown by colored dots, while the median $z$ as a function of $M_{\star}$ is highlighted by the thick lines in the main panel. This figure shows that at a given stellar mass, the ALFALFA sample is restricted to $z \lesssim 0.06$ compared to GAMA and DESI, which, on the other hand, show similar redshift distributions (see also vertical axis histograms). Faint galaxies in our sample are relatively nearby, while for MW-mass systems the volume surveyed is much larger. We have explicitly checked that the main results discussed in this paper do not show a significant evolution over the range of redshift studied here.

\begin{figure}
	\centering
	\includegraphics[width=\columnwidth]{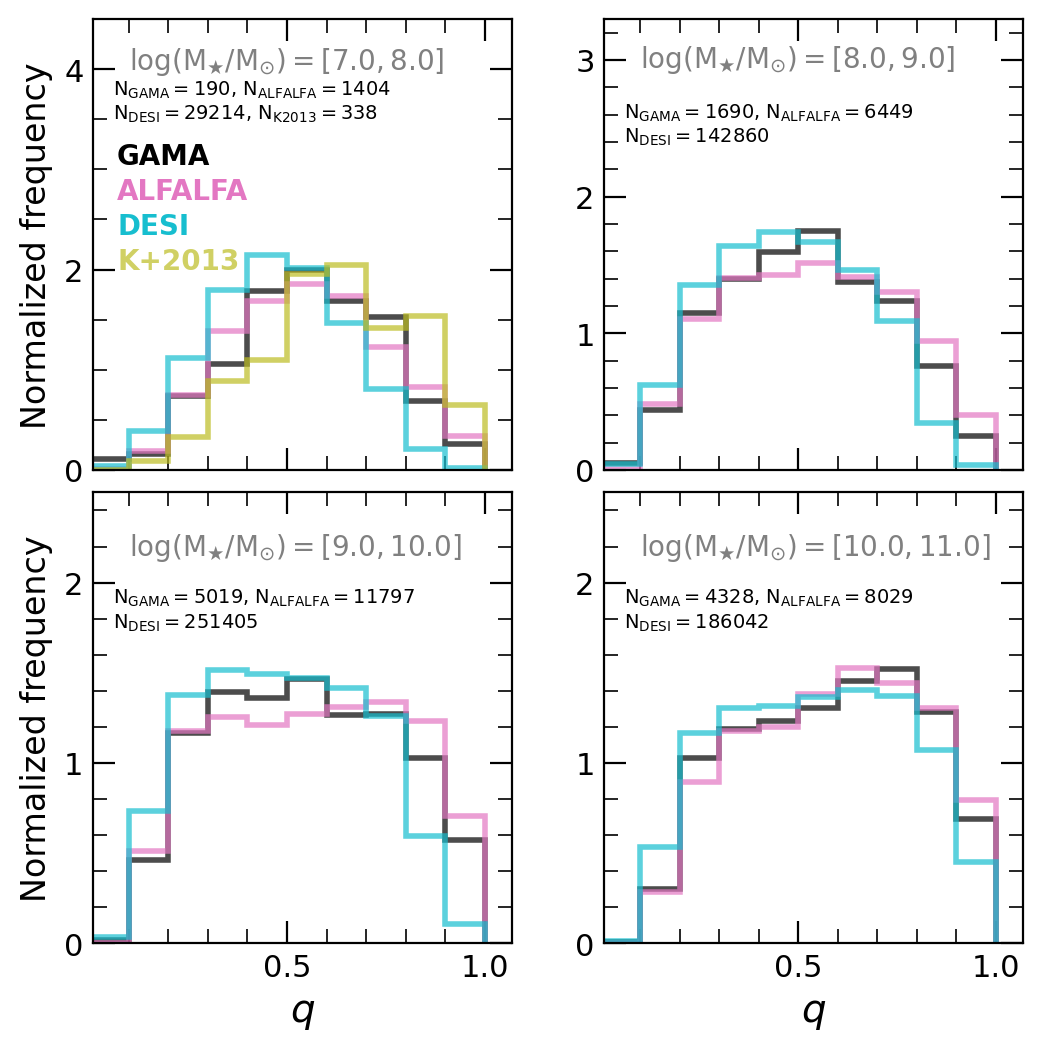}
	\caption{Normalized distributions of projected axis ratio ($q$) in bins of stellar mass. Different colors correspond to different observational surveys and the number of galaxies in each bin for each case is quoted in the panels. Galaxies with $M_{\star}>10^9 ~ \rm{M_{\odot}}$ show relatively flat $q$ distributions, but they become peaked at around $q\sim0.4 \rm - 0.5$ at lower masses. Notice that all stellar mass bins show at least some fraction of highly elongated galaxies with $q<0.3$.}
	\label{fig:distributions_q}
\end{figure}

\begin{figure*}
	\centering
	\includegraphics[width=1.0\textwidth]{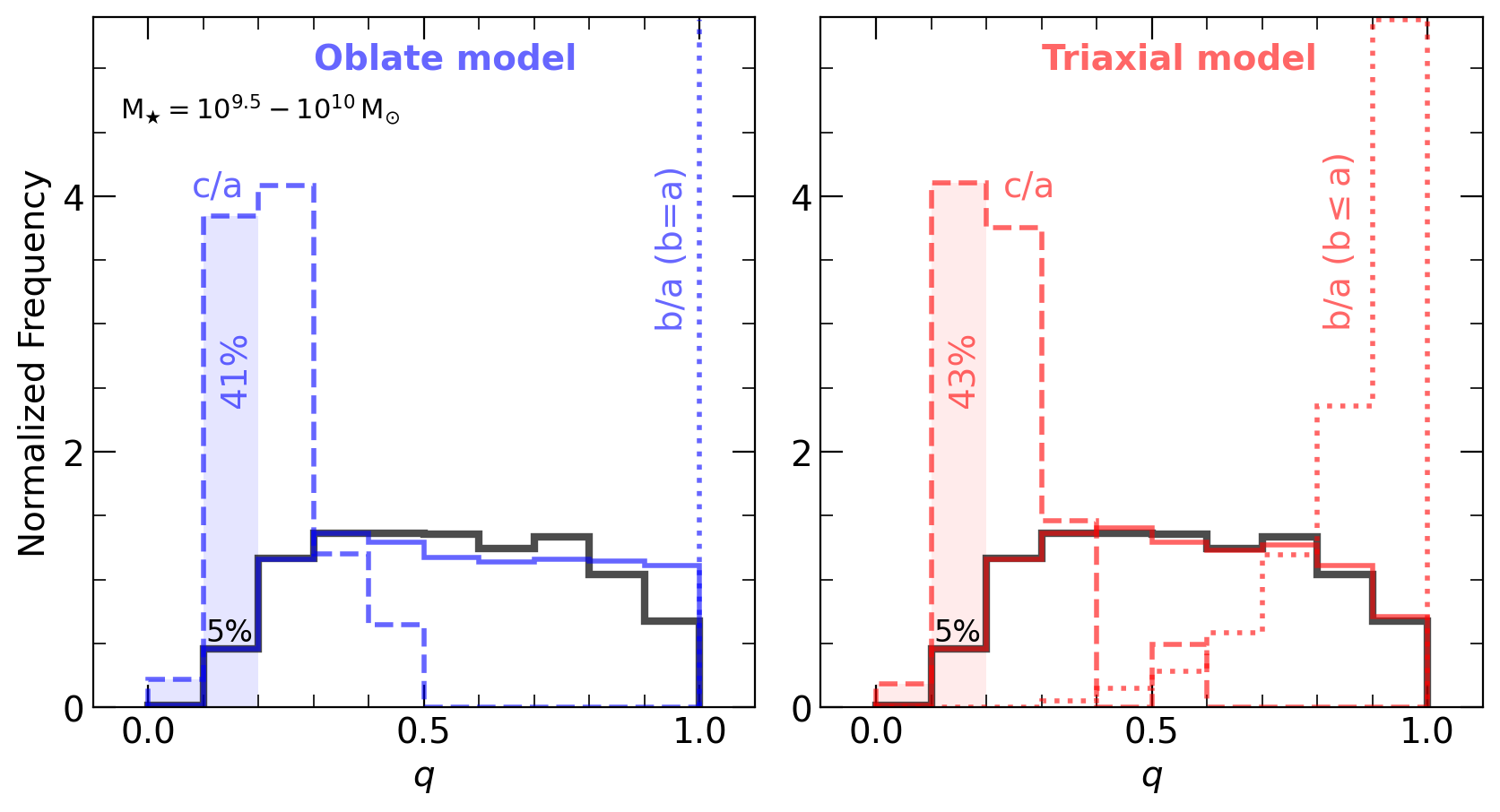}
	\caption{Comparison between the observed (projected) $q$ distribution and the inferred intrinsic 3D axis ratios for a subsample of dwarfs in the GAMA survey with $M_{\star}=10^{9.5} $-$ 10^{10} ~ \rm{M_{\odot}}$. The observed shapes are shown with the solid black line and are used to derive an intrinsic $c/a$ (long-dashed) and $b/a$ (dotted) distribution that the low-$q$ distribution when projected (method described in Sec.~\ref{SSecMockDistributions}). Solid color lines in each panel corresponds to the $q$ values obtained from the 3D model after random projections. Note that a wide distribution of $q$ values is generated by a much narrower intrinsic $c/a$ distribution. For example, the blue shaded region in the left panel shows that an observed $5\%$ of galaxies with $q<0.2$ requires $41\%$ of galaxies to have an intrinsic $c/a<0.2$ for an oblate model. Similarly, for a triaxal model (right panel, red curve) $43\%$ of galaxies are required to be thinner than $c/a=0.2$. The additional freedom of $b \neq a$ in the triaxial model helps to obtain a better fit to the projected $q$ distribution, but the changes mostly affect large $q$ values and changes little the $c/a$ frequency derived from highly elongated objects.}
	\label{fig:mock_example}
\end{figure*}

\subsection{Galaxy projected shape distributions}
\label{SSecShapes}

As discussed in Sec.~\ref{SecIntro}, the distribution of the projected axis ratio, $q=b_{\rm p}/a_{\rm p}$, provides a simple yet informative measure of the morphologies of observed and simulated galaxies, and, in particular, of the prevalence of disk and spheroidal components. Fig.~\ref{fig:stamps} showcases a small subset of DESI galaxies selected  to span the full stellar mass range of our study (in different rows) and $q$ values (columns), starting from highly elongated systems (left, $q < 0.2$) to nearly round objects (right column, $q \geq 0.8$). All images were generated using the DESI Legacy Survey database\footnote{\href{https://www.legacysurvey.org/}{https://www.legacysurvey.org/}} \citep{Zou2017, Dey2019, Moustakas2023, Myers2023}.

Galaxies in our smallest $q$ bin (leftmost column) are generally disc galaxies seen nearly edge-on,  while $q>0.8$ selects either face-on disks or spheroidal systems (rightmost column). Intermediate $q$ values (middle columns) display a wide variety of systems, from inclined discs to bars to flattened spheroids to tidally disturbed galaxies.

We study the $q$ distribution for galaxies in each of our samples after splitting them into four stellar mass bins, as shown in Fig.~\ref{fig:distributions_q}. The bins are one dex in width each and span the range $10^7<M_{\star}/\rm{M_\odot}<10^{11}$. Different colors correspond to different catalogs: GAMA (black), DESI (cyan) and ALFALFA (pink). In addition, we also show the results for  the Nearby Galaxy Catalog (olive) in the smallest mass bin, the only bin with enough systems in that catalog to allow for a robust measurement.

Despite the differences in sample selection and analysis techniques, there is remarkable agreement in the $q$ distributions of all three main surveys, at least in the three most massive bins. The two most massive bins have flat-topped distributions with few ultra-thin or very round objects. The $10^8$-$10^9 ~ \rm{M_\odot}$ bin is different, with an asymmetric $q$ distribution that peaks at $q=0.5$ and declines fairly rapidly towards flatter or rounder values. These results are  in good agreement with trends reported in earlier literature \citep[e.g., ][]{SanchezJanssen2010,Roychowdhury2013,KadoFong2020}.

The ``peaked'' distribution is also seen in the smallest-mass bin ($10^7$-$10^8 ~ \rm{M_\odot}$), where the agreement between surveys is less good, with the Nearby Galaxy Catalog (olive) being the most discrepant. The latter sample contains a larger fraction of high-$q$ systems, likely because many galaxies in this catalog are satellites of more massive hosts, where tides may have disturbed thin structures.

These distributions also make clear that highly elongated systems (say, $q<0.2$ or $0.3$) are present at all masses. As the examples in Fig.~\ref{fig:stamps} illustrate, these are almost invariably highly inclined thin stellar discs.  Disks may be less frequent, but do not ``vanish'' for dwarf galaxies, at least down to the lowest stellar masses probed here, $M_{\star} \sim 10^7 ~ \rm{M_\odot}$. 

\section{The galaxy intrinsic shape distribution}
\label{SecModel}

The {\it intrinsic} galaxy shape distribution may be derived from the {\it projected} $q$ distribution under simplifying assumptions; for example, that galaxies are mainly prolate or oblate axisymmetric figures of revolution \citep{Franx1991,Ryden1992}. Literature studies have long suggested that neither assumption is consistent with the observed shape distributions, and that some degree of ``triaxiality'' is required to reproduce the $q$ distributions shown in Fig.~\ref{fig:distributions_q} \citep[see, e.g,][]{Binggeli1980, Lambas1992, Padilla2008, Rodriguez2016, KadoFong2020, Rong2024}.

One reason for this is the fact that both prolate and oblate objects are round when seen ``face-on'' or ``end-on'', respectively, which implies that high-$q$ values should not be rare in either scenario. The sharp drop at high $q$-values seen in the distributions of  Fig.~\ref{fig:distributions_q} therefore cannot be reconciled with simple axisymmetric geometries \citep[e.g.,][]{BinneyVaucouleurs1981, BinneyMerrifield1998, Lambas1992}. Assuming that galaxies are triaxial objects (with three principal axes, $a>b>c$) solves this problem, and a number of studies have derived constraints on the $b/a$ and $c/a$ distributions consistent with the observed $q$ distributions \citep[e.g.,][]{Burkert2017, MendezAbreu2018, KadoFong2020}.

We address this issue here in a slightly different way. Rather than attempting to find the triaxial model distribution that best fits the data, we focus mainly on the highly-elongated objects which, as Fig.~\ref{fig:stamps} illustrates, are nearly always disks. This means that the oblate assumption most likely applies to these objects, which allows us to derive robust constraints on the prevalence of thin disks (strictly speaking, highly flattened objects) at given stellar mass. We are particularly interested in the faint galaxy end, both to determine the incidence of flat dwarfs, as well as to better characterize the transition from disk-dominated objects to spheroidal systems at the faint end.

\subsection{3D shapes of highly elongated galaxies}
\label{SSecMockDistributions}

To interpret the abundance of highly elongated systems, we first generate idealized 3D galaxy models by setting up $N$ ``stars'' that follow spatially a spherically-symmetric exponential radial density law. This spherical model can then be flattened along one or two orthogonal axes to reproduce  an arbitrary combination of axis ratios ($b/a$, $c/a$). These models can then be projected along a random direction, and their projected shapes (i.e., $q=b_{\rm p}/a_{\rm p}$) measured. (We have repeated this exercise varying the number of particles, and in general we observe that above $10,000$ particles, the projected shape measures are robust.)

For each projection of the model a $q$ ratio can be measured by either fitting ellipses to the projected isodensity contours (``isophotes'') or by computing the eigenvalue ratio of the 2D inertia tensor. The two methods give similar results, but the inertia tensor estimator performs better for the most elongated systems (i.e., the thinnest galaxies seen nearly edge-on), where the isophotal method tends to fail, especially for very thin objects seen nearly edge-on. We discuss this in more detail in Appendix~\ref{app:app2}.

These idealized models may be used to estimate the incidence of flat objects. If galaxies are oblate, then the fraction of highly-elongated galaxies gives a firm lower limit to the fraction of thin disks, simply because a galaxy with intrinsic $c/a$ equal to, say, $0.2$, cannot be have $q<0.2$ in projection, no matter the viewing angle.

To work out the fraction of intrinsically thin galaxies implied by the fraction of highly elongated galaxies in projection,  we work with the $q$ distribution observed in each stellar mass bin. Beginning with the lowest $q$ bin (i.e., $0 < q \leq 0.1$) we generate as many random projections of our idealized galaxy models (each with $b=a$ and $c/a$ chosen randomly between $0$ and $0.1$) as needed until we match the number of observed galaxies in that bin. The total number of trials is recorded, as well as the number that land in each $q$ bin, in this case the lowest bin and all bins at larger values. We then repeat the process for the next bin ($0.1 < q \leq 0.2$), generating again as many trials (with $0.1<c/a<0.2$) as needed to match the observed number of galaxies in that bin. These will contribute only to the second-lowest $q$ bin and bins at larger $q$ values. And so on.

The procedure is iterated over until the cumulative number of random trials equals the total number of galaxies in that stellar mass bin. This procedure ensures that the $q$ distribution is matched for low values of $q$, but there is no guarantee that the full $q$ distribution will be matched, especially at high values of $q$. The whole procedure can be repeated many times, to estimate uncertainties in the inferred $c/a$ distribution. We hereafter refer to the $c/a$ distributions inferred in this way as ``oblate''.

\subsection{The abundance of intrinsically thin GAMA galaxies}

We illustrate our results in the left panel of Fig.~\ref{fig:mock_example}, where we show the $q$ distribution for GAMA galaxies with $10^{9.5}<M_{\star}/\rm{M_\odot} < 10^{10}$ (thick black histogram). The blue solid line shows the match to the observed (projected) $q$ distribution that results from applying the outlined procedure using 10 equally spaced $q$ bins. As anticipated, the match is, by design, excellent at low $q$ values but starts to fail for $q>0.4$ because of our oblate assumption. The dashed blue curve shows the intrinsic $c/a$ distribution results in the blue solid histogram when viewed in projection.

The most notable result of this exercise is that even a relatively small fraction of highly-elongated galaxies implies a substantial fraction of thin disks.  For instance, only $5\%$ of the galaxies in the observed $q$ distribution have $q<0.2$, but this implies that roughly $41\%$ of the sample should be made of disks thinner than  $c/a<0.2$ (shaded area). To reproduce the observed $16\%$ of flat objects with $q<0.3$, nearly $\sim 80\%$ of the galaxies should have intrinsic shapes flatter than $c/a=0.3$. Clearly, thin stellar disks make up the majority of galaxies in this stellar mass bin.

One may worry that this conclusion is highly dependent on our oblate assumption, which leads to projected distributions that deviate from the observed ones at high $q$ values. We therefore consider the possibility of ``triaxial'' systems  by varying the intrinsic $b/a$ ratio of the idealized galaxy models. To keep the procedure simple, we add a single extra adjustable parameter, $\alpha$, so that the $b/a$ cumulative distribution is given by: 
\begin{equation}
f(b/a) = \frac{1 - e^{(b/a)^{1/\alpha}}}{1 - e}.
\label{eq:ba_func}	
\end{equation}

Here, $\alpha \sim 0$ corresponds to the case $b=a$, and increasing values of $\alpha$ generate more ``triaxiality''. We then repeat the $q$-fitting procedure outlined above, varying $\alpha$ until the model and observed $q$ distributions are best matched. In practice, this means the value of $\alpha$ which minimizes $\sum_{i} (N_{\rm{ model_i}} - N_{\rm{ data_i}})^2  / N_{\rm{ data_i}}$.

This triaxial model leads to much better agreement between predicted and observed $q$ distributions, as seen in the right-hand panel of  Fig.~\ref{fig:mock_example}. The dotted red line in this panel shows the $b/a$ distribution of $N_{\rm gal}$, obtained from Eq.~\ref{eq:ba_func} for $\alpha=0.2$. Clearly, most systems are nearly oblate; $\sim 54 \%$ of them have $b/a>0.9$. More importantly, the $c/a$ (dashed red curve) distribution has hardly changed, especially at low $q$. For the best fit model, this new procedure implies an slightly higher fraction of thin ``disks'', with $43\%$ and $80\%$ thinner than $c/a=0.2$ and $0.3$, respectively.

\begin{figure}
	\centering
	\includegraphics[width=\columnwidth]{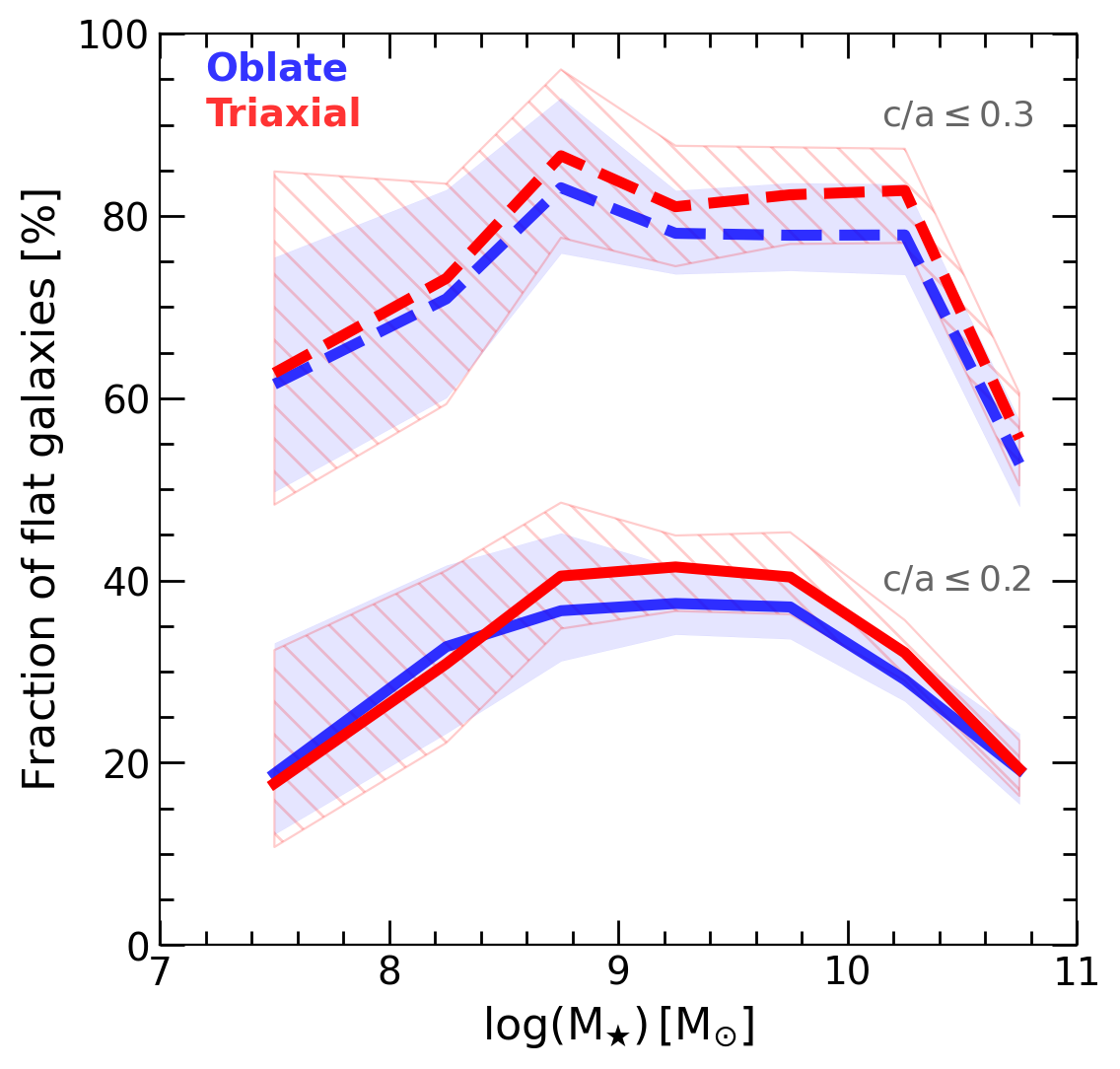}
	\caption{Fraction of galaxies in the GAMA sample that are derived to be intrinsically thinner than a given $c/a$ cut as a function of stellar mass. We show $c/a<0.2$ and $0.3$ in thick solid and dashed lines, respectively. Shaded regions correspond to the $10^{\rm th}$-$90^{\rm th}$ percentiles from repeating the intrinsic shape reconstruction $100$ times in each $M_{\star}$ bin (see Sec.~\ref{SSecMockDistributions} for details). The fraction of intrinsically thin galaxies peaks for galaxies with $M_{\star}=10^9$- $10^{10}\, \rm{M_{\odot}}$. Following the color scheme in Fig~\ref{fig:mock_example}, values for the oblate and triaxial models are shown in blue and red, respectively, but estimates of the fraction of thin galaxies are consistent in both models.}
	\label{fig:oblate_triaxial}
\end{figure}

\begin{figure*}
	\centering
	\includegraphics[width=1.0\textwidth]{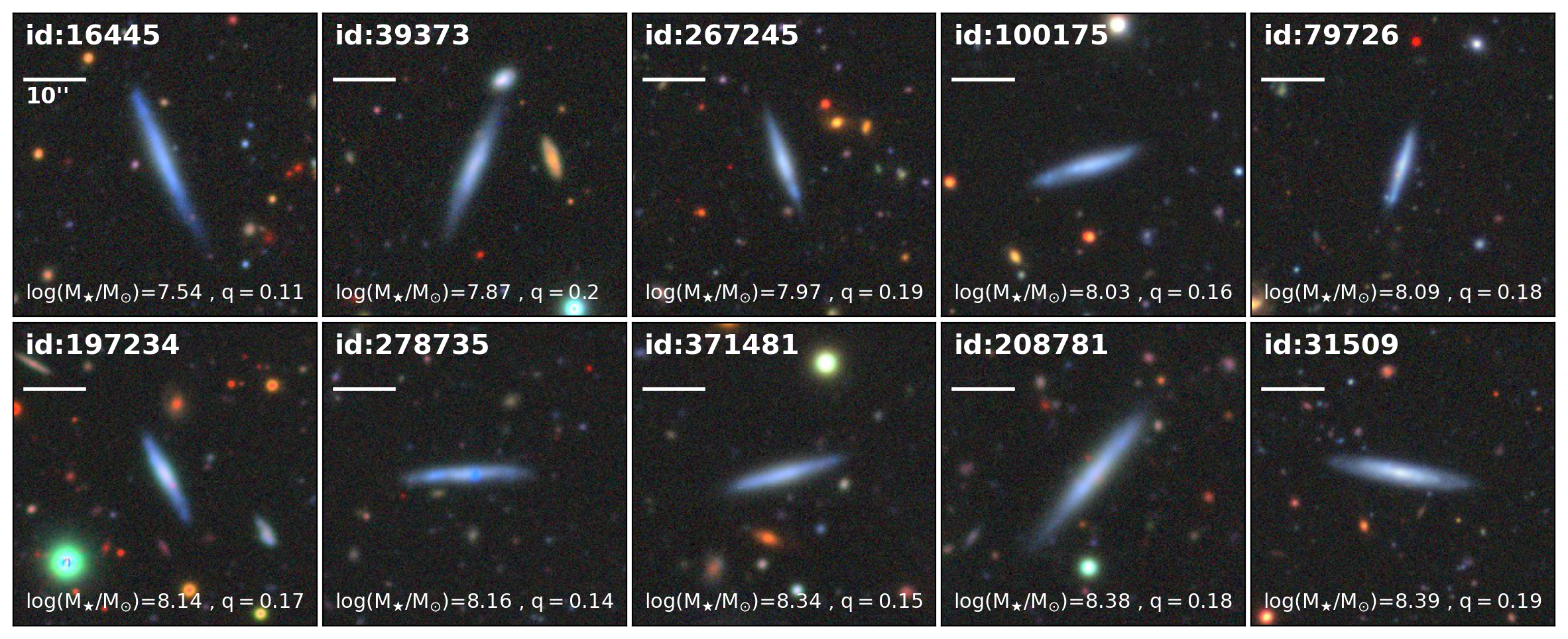}
	\caption{Examples of $10$ highly elongated dwarf galaxies with $q \leq 0.2$ and $M_{\star}=10^7$ - $10^{8.5} ~ \rm{M_{\odot}}$. They resemble thin edge-on disks and can be found even among the faintest dwarfs in our sample. Legends in each panel quote the stellar mass, the shape parameter $q$, as well as the GAMA identifier. Objects are sorted by increasing $M_{\star}$, left to right. All images were generated using the Legacy Survey database.}
	\label{fig:examples_low_q}
\end{figure*}

\begin{figure*}
	\centering
	\includegraphics[width=\columnwidth]{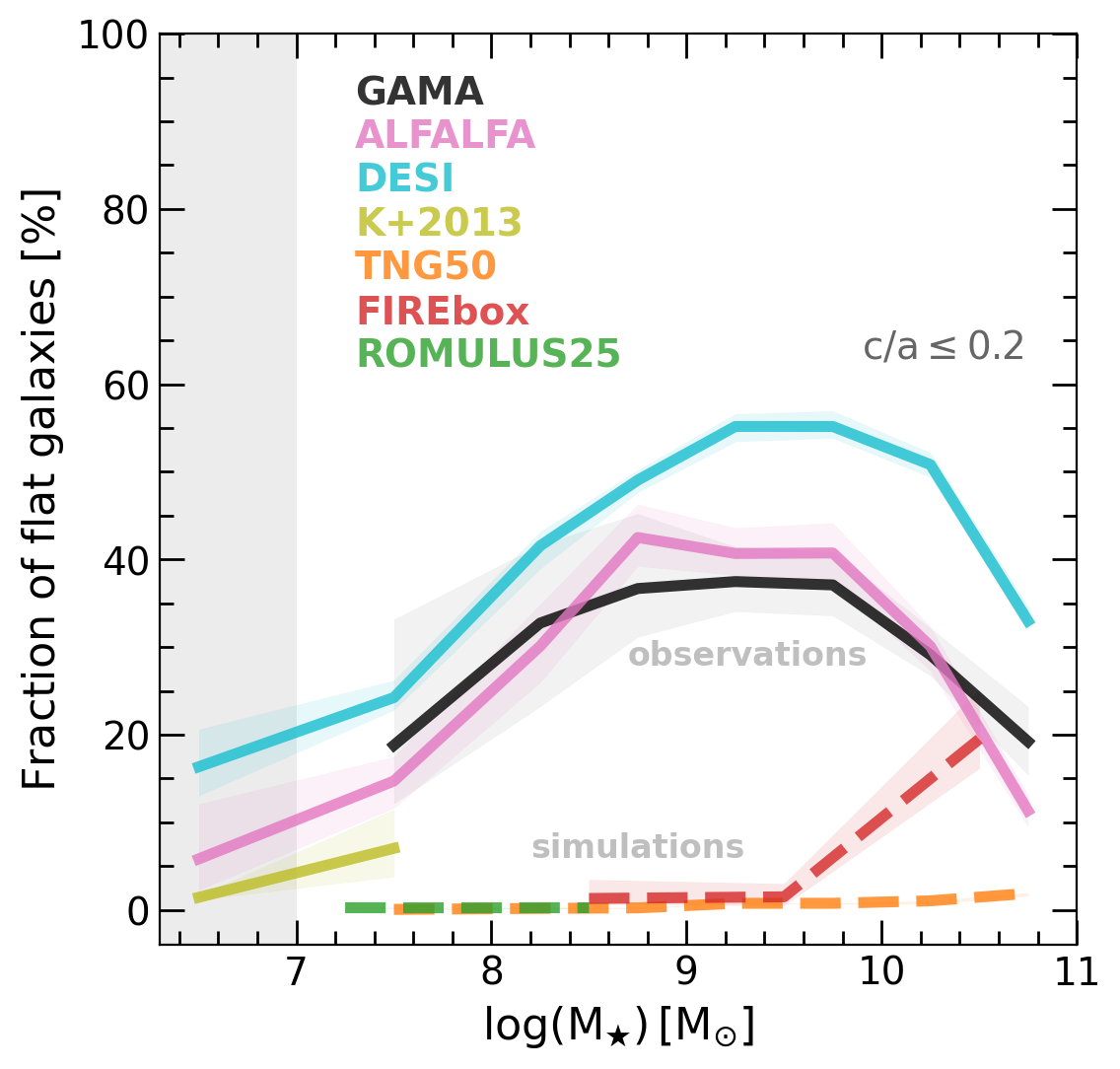}
	\includegraphics[width=\columnwidth]{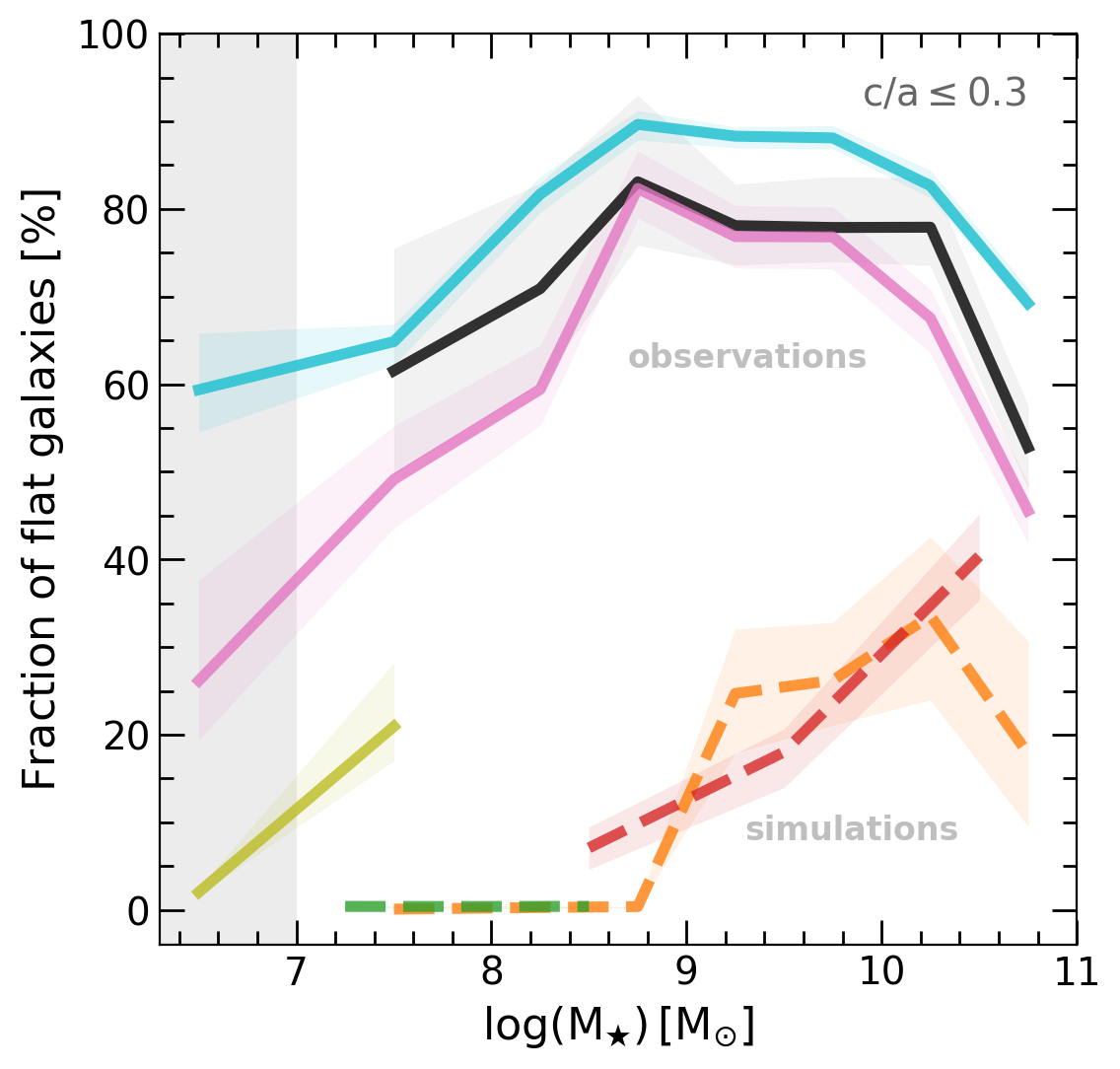}
	\caption{Fraction of galaxies that are derived to be intrinsically thinner than $c/a \leq 0.2$ (left) and $c/a \leq 0.3$ (right) as a function of stellar mass. Thick solid lines correspond to our observational samples while dashed lines are used to display the results of cosmological simulations. Different colors highlight the specific survey or simulation name, as quoted in the legend. In all observational surveys, the frequency of thin galaxies peaks for dwarfs with $M_{\star}\sim 10^9 ~ \rm{M_{\odot}}$, almost doubling the frequency observed on the scale of MW-mass galaxies. Thin galaxies do not disappear at lower masses: we infer a significant fraction of dwarf galaxies with $M_{\star}<10^9 ~ \rm{M_\odot}$ to have $c/a<0.2$. This is in stark contrast with the negligible production of thin dwarf galaxies in all numerical simulations analyzed here.}
	\label{fig:percent}
\end{figure*}

We apply the procedure outlined above to our full GAMA sample to study the minimum fraction of thin disk-like objects needed to reproduce the observed distribution of highly elongated galaxies as a function of stellar mass. The results are shown in Fig.~\ref{fig:oblate_triaxial}, where solid and dashed lines indicate the fraction of systems with $c/a<0.2$ and $0.3$, respectively. To estimate uncertainties, we have repeated the procedure $100$ times for each mass bin. The $10^{\rm th}$-$90^{\rm th}$ percentiles are shown as shaded regions around each curve. We have explicitly checked that our recovered fraction of thin galaxies remains statistically consistent when allowing for Gaussian uncertainties in $q$  with $\sigma_q=0.05$ width. This is a conservative estimate, as the average error in $q$ is $0.01$ in our GAMA sample. 

The oblate and triaxial models are shown in Fig.~\ref{fig:oblate_triaxial} in blue and red colors, respectively, and are in good agreement with each other, confirming that, despite its simplicity,  the oblate assumption provides robust estimates of the incidence of thin disks in GAMA.

Fig.~\ref{fig:oblate_triaxial} shows that in general galaxies are intrinsically quite flat: thin disks with $c/a < 0.2$ make up $20\%$ to $40\%$ (solid) of all galaxies with stellar masses $10^7<M_{\star}/\rm{M_\odot}<10^{11}$. These fractions climb to $60\%$ to $80\%$ for slightly thicker, but still quite flattened $c/a < 0.3$ galaxies (dashed).

More importantly, the incidence of thin disks peaks in the range $M_{\star}=[10^9$-$ 10^{10}] ~ \rm{M_\odot}$, and decreases towards both less and more massive objects. Galaxies with a mass comparable to the Milky Way ($\sim 5\times 10^{10}~ \rm{M_\odot}$), in other words, are not the archetypal thin disk galaxy in the Universe; rather, galaxies $10$ to $50$ times less massive, a range that includes M33 and the LMC, for example, are those where thin disks are most prevalent.

The incidence of thin disks declines for $M_{\star}<10^9~ \rm{M_\odot}$, but only gradually; even for our smallest mass bin ($M_{\star}=[10^7$-$10^8]~ \rm{M_\odot}$), galaxies with  $c/a < 0.2$ make up at least $20$-$25\%$ of our sample. This illustrates one of the main findings of this paper: thin dwarf galaxy disks are not rare, a result with important corollaries for present-day cosmological hydrodynamical simulations, as we discuss  below.  We showcase in  Fig.~\ref{fig:examples_low_q} some of these highly elongated dwarfs in the range $M_{\star}=[10^7 \rm - 10^{8.5}]~ \rm{M_\odot}$. These images nicely illustrate the presence of well defined, thin stellar disks in low-mass galaxies. 

\begin{figure}
	\centering
	\includegraphics[width=\columnwidth]{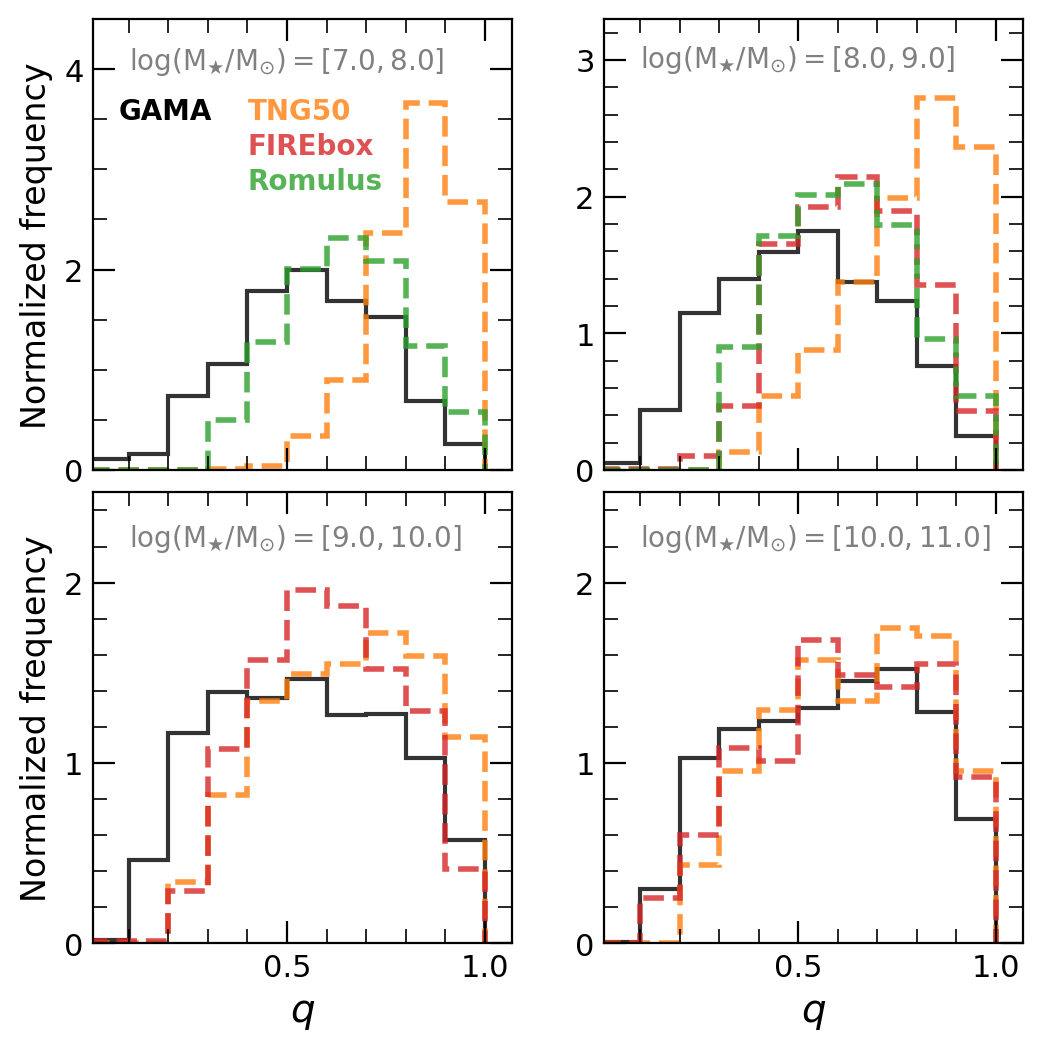}
	\caption{Normalized distribution of projected axis ratio ($q$) in stellar mass bins for our simulated galaxy samples: TNG50 (orange), FIREbox (red), Romulus (green). For reference we include the observed $q$ distribution from the GAMA sample, as shown in Fig.~\ref{fig:distributions_q}. All numerical simulations severely underproduce thin dwarfs with $q<0.3$. The agreement between observed and simulated galaxy shapes improves somewhat on the scale of MW-mass galaxies  (bottom right panel).} 
	\label{fig:q_sims}
\end{figure}

\vspace{-0.1cm}

\subsection{Thin disks in DESI and ALFALFA}
\label{SecSimulations}

Fig.~\ref{fig:percent} extends our analysis to other galaxy samples (solid lines), and shows that the main trends highlighted in Fig.~\ref{fig:oblate_triaxial} are also found in other surveys, such as DESI and ALFALFA. Given the good agreement shown in GAMA between the oblate and triaxial modeling, in what follows we report results only for the oblate model. The results for DESI, in particular, are interesting because it is the most numerous sample, and also because it is based on the latest photometric measurements. Encouragingly, samples with very different selection effects, such as the optically-selected GAMA and HI-selected ALFALFA galaxies, return similar estimates for the fraction of flat galaxies, strengthening our results and indicating that it is unlikely born of optical selection biases.

Notably, the DESI results indicate that the incidence of thin disks inferred from GAMA may actually be an underestimate of the true abundance. Indeed, the DESI fractions exceed even those of ALFALFA, which, because of its HI selection, would be expected to favor gaseous and stellar disks. The exact origin of this systematic discrepancy is unclear (see Appendix~\ref{app:app1} for a discussion), but suggests that the fraction of thin galaxies inferred from GAMA and ALFALFA are conservative lower limits to the true abundance of thin disk galaxies at given stellar mass. 

Note also that we have extended the lower mass bins in Fig.~\ref{fig:percent} by another dex in order to accommodate the results from the Nearby Galaxy Catalog. The fraction of thin disks in this catalog is lower than in other samples, as discussed in Sec.~\ref{SSecShapes}.

\vspace{-0.05cm}

\subsection{Thin disks in cosmological simulations}

The dashed thick lines in Fig.~\ref{fig:percent} present a compilation of results on galaxy shapes from cosmological hydrodynamical simulations in the literature, including TNG50 (orange), FIREbox (red) and ROMULUS25 (green). Sec.~\ref{SSecSims} briefly describes each simulation and how the information on galaxy shapes has been extracted. In all cases, we start from the projected $q$ distributions of simulated galaxies, which are shown in Fig.~\ref{fig:q_sims}. We then apply the same procedure described in Sec.~\ref{SSecMockDistributions} for observational samples to derive the fraction of thin simulated galaxies shown in Fig.~\ref{fig:percent}.

The comparison of simulations with observations in Fig.~\ref{fig:percent} and Fig.~\ref{fig:q_sims} implies a clear deficiency of thin simulated galaxies with $M_{\star}< 10^{10}~ \rm{M_\odot}$. This deficiency applies to {\it all} analyzed simulations. The disagreement between simulations and observations peaks in the regime of dwarf galaxies with $M_{\star}<10^9~ \rm{M_\odot}$. Only the FIREbox simulation reproduces the frequency of thin objects ($c/a<0.2$), but in the most massive bin, corresponding to MW-mass galaxies.

The formation of a large number of disk galaxies is an undeniable success of the TNG50 model \citep{Pillepich2019}. In particular, these authors highlight that MW-mass simulated galaxies are quite thin, using as a metric the ratio between the half-height of disk stars to the cylindrical half-mass radius, $h_{1/2}/r_{h,*}<0.2$. This metric, unfortunately, is not accessible to observations, and quite biased towards the inner regions, often dominated by a bulge. Therefore, alternative shape estimates, such as $q$ or the inferred $c/a$ using the inertia tensor, as done here, allow for a fairer comparison between observations and simulations.

Turning our attention back to dwarf galaxies, the upper two rows in Fig.~\ref{fig:q_sims} show that this discrepancy in the low-mass regime is also clear when the projected shapes $q$ are compared: there are essentially no simulated dwarfs (i.e., $M_{\star}<10^9~ \rm{M_\odot}$) with $q<0.3$,  in clear contradiction with the GAMA results shown by the gray solid histogram. A similar conclusion has also been reported by \citet{Klein2025} for the FIREbox simulations. The red dashed curves in  Fig.~\ref{fig:q_sims} confirm their conclusion. TNG50 dwarfs (orange dashed lines) are even more spheroidal, with a $q$ distribution that is remarkably different from the observed one. More recently, \citet{Dado2026} have reported dwarf galaxy shapes for the COLIBRE simulation \citep{Schaye2025}; these also seem inconsistent with the high fraction of flattened galaxies inferred from the galaxy surveys used in this work. 

We have explicitly checked that our results are robust to two factors that could influence shape measurements: $i)$ the aperture within which $q$ is determined and $ii)$ the fraction of quenched systems in the sample. \citet{KadoFong2020} have shown that measured shapes become more spheroidal with radius when $q$ is measured as far as $4$ effective radii. Using our TNG50 sample, we tested the impact of this apperture dependence on our reported fraction of thin galaxies, finding a weak dependence that is insufficient to reconcile simulations with observations.Similarly, the disagreement between observational data and the results reported for FIREbox by \citet{Klein2025} are not the result of apperture effects, as these authors measured shapes using SDSS mock images, which naturally reproduce surface brightness limits consistent with those in our sample. Put together, these tests suggests that aperture or surface brightness limits are not the main culprit of the discrepancy between simulations and observations we report here.

In addition, the disagreement with observed shapes persists even when restricting the simulated galaxies in TNG50 to only star-forming galaxies. Such selection should weaken any secondary effects on the shape distribution driven by a potentially higher-than-observed fraction of quenched galaxies in isolation \citep{KadoFong2025}.

\begin{figure*}
	\centering
	\includegraphics[width=1.0\textwidth]{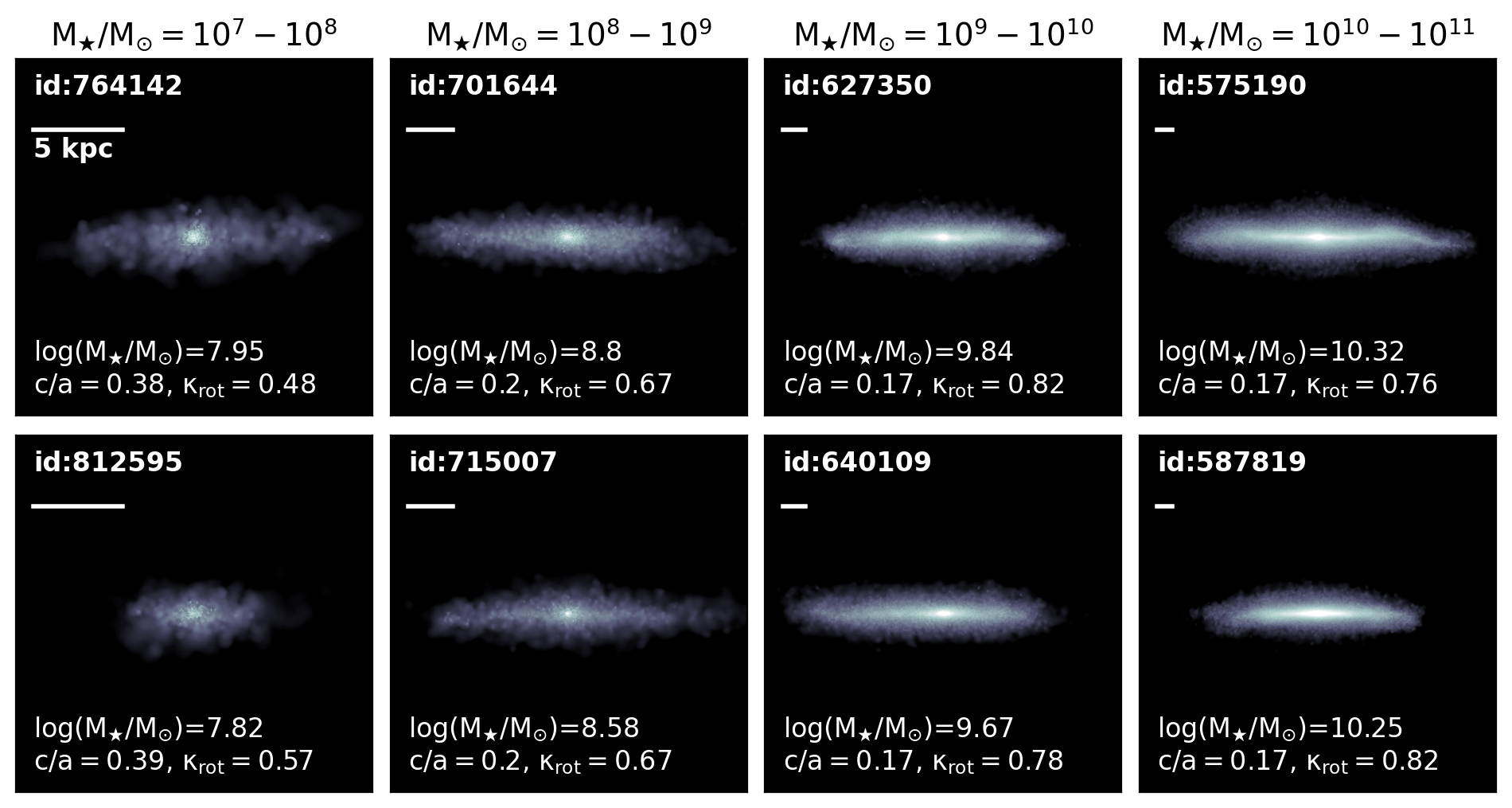}
	\caption{Edge-on projection of examples of the flattest galaxies in the TNG50 simulation, in different bins of stellar mass. The rotation is determined by the angular momentum of the galaxy disk. Images were generated using the \textsc{Py-SPHViewer} code \citep{BenitezLlambay2017}.}
	\label{fig:expls_lowq_TNG}
\end{figure*}

Fig.~\ref{fig:expls_lowq_TNG} presents some edge-on examples of the flattest dwarf galaxies in the TNG50 simulation, rotated in such a way that the angular momentum points in the vertical direction (perpendicular to the plane of the galaxy). We have chosen the galaxies with the lowest $c/a$ in each bin of stellar mass. The $\kappa_{\rm rot}$ parameter \citep{Sales2012} of each galaxy, which defines the fraction of kinetic energy in rotational support, is included in each panel, indicating that most of these flat galaxies are rotationally supported. We want to highlight the fact that, at least for $M_{\star} > 10^8 ~ \rm{M_\odot}$, some simulated galaxies as flat as $q\sim 0.2$ are actually found in these state-of-the-art cosmological simulation, but not nearly as many as needed to reproduce the  observational results. We discuss several caveats regarding the effects of numerical resolution in App.~\ref{app:resolution}.

\vspace{-0.1cm}

\section{Discussion}
\label{SecDiscussion}

The lack of thin dwarfs  in cosmological simulations is not a subtle problem. While observations indicate that $80\%$ of dwarfs with $M_{\star} \sim 10^9 ~ \rm{M_\odot}$ are thinner than $c/a=0.3$ (we shall quote quantitative results  in this section for GAMA, which likely underestimate the true frequency of flattened objects, see Sec.~\ref{SecSimulations}), and about half of them thinner than $c/a=0.2$ (Fig.~\ref{fig:percent}), simulations produce fewer than $\sim 20\%$ of dwarfs of this mass with $c/a<0.3$ and {\it none} thinner than $c/a=0.2$. The disagreement grows worse below $M_{\star} \sim 10^9~ \rm{M_\odot}$, where all simulated galaxies have $c/a>0.3$, while the observed abundance of galaxies flatter than $c/a=0.3$ varies from a whopping $\sim 80\%$ for LMC-mass dwarfs ($10^9 ~ \rm{M_\odot}$) to $\sim 60\%$ in our smallest mass bin of $10^7<M_{\star}/\rm{M_\odot} <10^8$.

Disks thinner than $c/a=0.3$ are abundant in low mass galaxies and fully dominate the population at the scale of LMC-like dwarfs. This is clearly not the case in simulations. This  statement relies on our assumption that highly elongated galaxies (i.e., low $q$) have disk-like structures. While full confirmation of this is not possible without kinematical data, several factors support this assumption. First, we have visually inspected many random subsamples of galaxies with low $q$ values, finding that in the vast majority of cases their morphology is consistent with edge-on disks and not bar-like galaxies or tidal features, at least for $q<0.3$. We include several examples of observed galaxies with low-$q$ values in our lowest mass bins in Fig.~\ref{fig:examples_low_q}, which clearly suggest a disk-like nature. Second, we have also selected low $q$ dwarfs from the ALFALFA sample, and retrieved their HI line profiles from the data presented in \citet{Haynes2018}. The low-$q$ objects inspected had broad HI line profiles consistent with that expected from (gas) rotation, albeit with moderate signal-to-noise. 

Therefore, while bars, for example, may contribute some flat low mass galaxies \citep[e.g.,][]{Wang2025}, they do not dominate the population of highly elongated galaxies. This is also supported by work in the literature that argues that bars are less common below $M_{\star} \sim 10^9~ \rm{M_\odot}$ and, even if they exist, they still inhabit dynamically cold disks \citep{MendezAbreu2010}.

Several other factors indicate that our comparison between observed and simulated galaxy shapes is conservative. For instance, only central galaxies have been included in the simulated samples, while observations include all environments. By selecting central galaxies in the simulations we have preferentially removed tidally-shaped objects and satellites, which tend to have more spheroidal shapes \citep[e.g., ][]{Dressler1980,Postman1984,Valluri1993, Lokas2020,Pfeffer2023}. In addition, observational estimates are affected by the presence of dust, which is not accurately modeled in simulations. It is well established that the presence of interstellar dust can significantly distort the determination of a galaxy shapes ($q$), especially when observed at optical wavelengths \citep{Disney1989, Maller2009}. There is evidence for this in observations, for example, \citet{Wild2011} shows how galaxies from the SDSS with highly inclined disks look optically thicker due to the attenuation of light from young stars due to diffuse dust in the ISM. And simulation-based studies such as \citet{Byun2025} report that the presence of dust in simulated galaxies leads to brighter centers and more pronounced bulges, affecting their intrinsic shapes and yielding a lower fraction of late-type systems. The lack of elongated dwarfs in the simulations despite this optimistic comparison highlights a clear disagreement with observations.

Our results also indicate that dwarfs become more spheroidal below stellar masses of a few times $10^8 ~ \rm{M_{\odot}}$, a result that echoes earlier observational work \citep{SanchezJanssen2010,Roychowdhury2013,KadoFong2020}. Some numerical simulations have also reported a similar trend, either by analyzing galaxy shapes or the degree of rotational support \citep{ElBadry2016, Zeng2024,Celiz2025,Benavides2025b,Keith2025}. Here we highlight that despite this trend with stellar mass, a substantial fraction of observed galaxies are still thin, even at the lowest mass bins that we analyzed. On the other hand, simulated dwarfs are all intrinsically too thick, with no disks with $c/a<0.3$ for $M_{\star}<10^9~ \rm{M_\odot}$, in clear contrast with observations.  This is in agreement with \citet{Klein2025}, who pointed out that the shape of FIREbox dwarfs was inconsistent with those in GAMA. We have here confirmed and extended that conclusion by including data from additional surveys like ALFALFA and DESI and by inferring their intrinsic shapes. The problem does not apply solely to FIREbox, but also to TNG50 and Romulus25, the other two simulations that we have included in this study.  

The observed abundance of thin dwarfs is also somewhat surprising given our current understanding of how galaxy formation proceeds in low mass systems. For example, the presence of thin stellar disks would seem to preclude large potential fluctuations during the late formation history of a dwarf. These fluctuations are, however, critical for transforming cold dark matter ``cusps'' into constant-density ``cores'', a transformation that has been advocated repeatedly in the simulation literature \citep[see, e.g., ][]{Navarro1996b,Read_and_Gilmore2005,Governato2012,DiCintio2014,Onorbe2015,Tollet2016,BenitezLlambay2019}. The efficiency of this process has been claimed to peak in the range $M_{\star} \sim 10^8$-$10^{9.5} ~ \rm{M_\odot}$, just where our results indicate that thin stellar disks are plentiful: $\sim 30$-$40\%$ have disks as thin as $c/a <0.2$ in this mass range \citep{Governato2012,DiCintio2014,Onorbe2015,Emami2019,Jackson2025}.

\section{Summary and Conclusions}
\label{SecConc}

We have studied the shape of galaxies as a function of stellar mass in the range $10^7<M_{\star}/\rm{M_\odot}< 10^{11}\, M_\odot$ in observations and simulations. We combine data from the observational surveys GAMA, DESI, and ALFALFA, as well as a small  sample of low-mass galaxies from the Nearby Galaxy Survey, to measure the distribution of galaxy projected shapes ($q$) as a function of mass. Galaxies with $M_{\star}>10^{9} ~ \rm{M_\odot}$ have a relatively flat $q$ distribution in the range $0.3<q<0.8$, whereas dwarfs ($M_{\star} < 10^{9} ~ \rm{M_\odot}$) have peaked distributions centered around $q \sim 0.4$. 

We have visually examined large numbers of highly-elongated objects and conclude that most of them consist of highly-inclined thin stellar disks. We use this to derive constraints on the abundance of intrinsically thin disks required to  match the observed frequency of highly-elongated objects.  We find that the fraction of thin galaxies with $c/a < 0.2$ peaks at $\sim 40\%$ for galaxies in the range $10^9<M_{\star}/\rm{M_\odot}< 10^{10}$, and decreases towards both fainter and more massive galaxies. The fraction of thin dwarfs with mass comparable to the SMC ($M_{\star} \sim 10^8 ~ \rm{M_\odot}$) is roughly equal to that of MW-like galaxies.

We focus our analysis on the scale of dwarf galaxies, where many cosmological hydrodynamical simulations predict that thin stellar disks should become rare. Observed dwarfs are generally quite thin: $20$-$35\%$ of dwarfs with $M_{\star}=10^8$-$10^9 ~ \rm{M_\odot}$ have intrinsic axis ratios $c/a<0.2$, and the fraction grows to  nearly $80\%$ for $c/a<0.3$. Even in our lowest galaxy mass bin ($10^7<M_{\star}/\rm{M_\odot}< 10^8$), $10$ to $20\%$ of dwarfs are thinner than $c/a =0.2$, a clear demonstration that thin stellar disks are able to form and survive in significant numbers in low mass dwarfs. 

Numerical simulations fail to satisfy these constraints, producing structures that are substantially more spheroidal than observed for galaxies with  $M_{\star} \leq 10^{10}~ \rm{M_{\odot}}$. This applies to the three state-of-the-art cosmological simulations that we included in our analysis: TNG50, FIREbox and systems with $M_{\star} < 10^9 ~ \rm{M_{\odot}}$ from Romulus25. No simulation is apparently able to form galaxies thinner than $c/a=0.2$ in systems with $M_{\star} \sim 3 \times 10^{8} ~ \rm{M_{\odot}}$, a mass range where GAMA observations suggest that between $\sim 27\%$ and  $ \sim 43\%$ of systems are at least that thin.

The same problem applies to more massive galaxies, which are better resolved in the cosmological simulations. In the mass range $M_{\star} = 10^9$-$10^{10} ~ \rm{M_{\odot}}$, where observations suggest that the fraction of thin galaxies with $c/a<0.2$ peaks (at $\sim 40\%$), we find that {\it all} TNG50 and FIREbox galaxies are thicker than $c/a = 0.2$.

The lack of thin stellar disks in simulated dwarfs has interesting implications. For example, it shows that, despite enormous improvements to our galaxy formation models in the last couple of decades, simulations are still unable to reproduce the population of galaxy disks on mass scales below that of the MW, suggesting that our understanding of the formation of galaxy disks remains incomplete. In addition, our results support the idea that the coupling of stellar feedback to the interstellar medium in current numerical models may be too strong, hindering the formation and survival of thin, rotationally-supported stellar disks. This is particularly puzzling in the $10^8<M_{\star}/\rm{M_\odot}< 10^9$ range, where many simulations have consistently suggested that feedback is most efficient, and able to drive substantial fluctuations in the gravitational potential that may even transform dark matter cusps into cores. How to square the substantial fraction of thin disks observed in dwarfs with the massive inflows and outflows of gas that simulations predict is a pressing challenge that future simulation work should attempt to solve.

\section*{Acknowledgments}
JAB and LVS are grateful for partial financial support from NSF-CAREER-1945310, NSF-AST-2107993 and NSF-AST-2408339 grants. This research was supported in part by grant NSF PHY-2309135 to the Kavli Institute for Theoretical Physics (KITP). Computations were performed using the computer clusters and data storage resources of the HPCC, which were funded by grants from NSF (MRI-2215705, MRI-1429826) and NIH (1S10OD016290-01A1). LVS would like to thank the Max Planck Institute for Astrophysics and KITP for hospitality and financial support during the completion of this work. JFN acknowledges the hospitality of the Max-Planck Institute for Astrophysics, the Donostia International Physics Center, and Durham University during this study. KAO acknowledges support by the Royal Society through Dorothy Hodgkin Fellowship DHF/R1/231105. This work used the DiRAC@Durham facility managed by the Institute for Computational Cosmology on behalf of the STFC DiRAC HPC Facility (www.dirac.ac.uk). The equipment was funded by BEIS capital funding via STFC capital grants ST/K00042X/1, ST/P002293/1,
ST/R002371/1, and ST/S002502/1, Durham University and STFC operations grant ST/R000832/1. DiRAC is part of the National e-Infrastructure.

\section*{Data Availability}

This paper is based on halo catalogs and snapshots from the Illustris-TNG Project \citep{Nelson2019TNG, Nelson2019TNG50}. These data are publicly available at \href{https://www.tng-project.org/}{https://www.tng-project.org/}. The main properties of the dwarf galaxy samples, and other products included in this analysis, may be shared upon request to the corresponding author if no further conflict exists with ongoing projects. 



\bibliographystyle{mnras}
\bibliography{biblio.bib} 



\appendix

\section{Comparison between DESI and GAMA shapes}
\label{app:app1}

\begin{figure}
	\centering
	\includegraphics[width=0.99\textwidth]{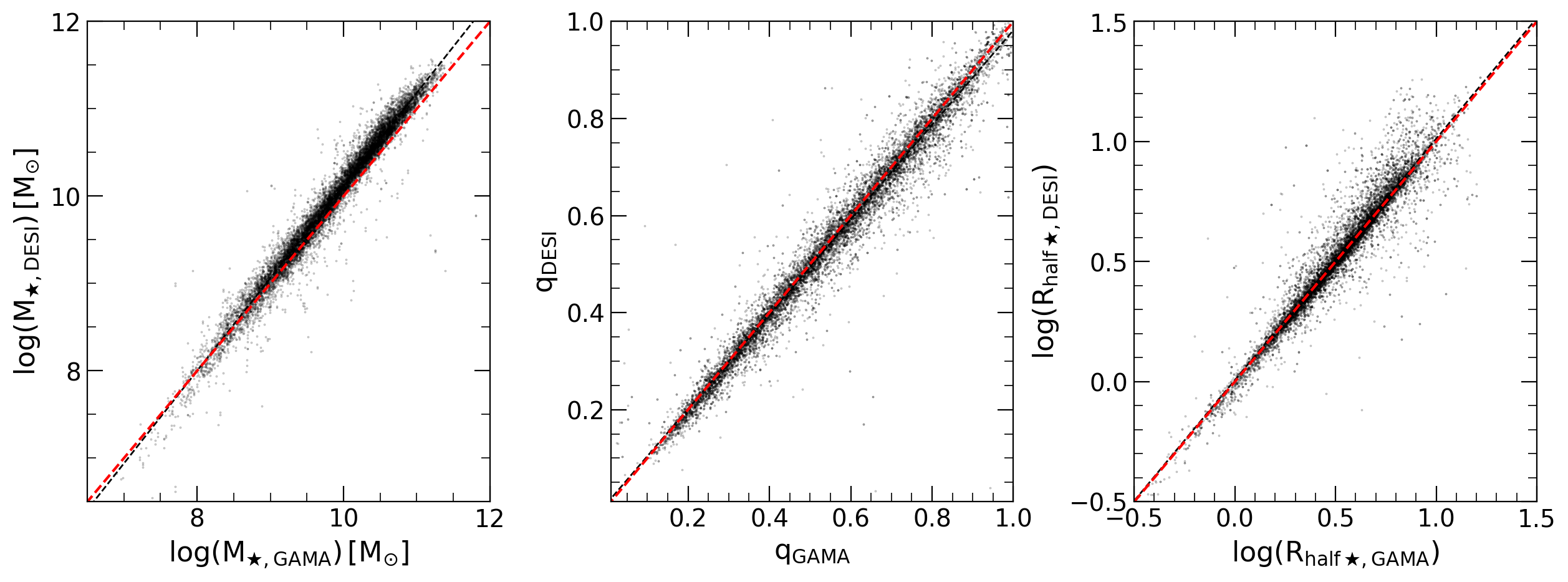}
	\caption{Comparison of stellar mass (left), shape parameter $q$ (middle), and half-mass radius, $R_{\rm{half , \star}}$ (right) for a sample of objects in common in the DESI and GAMA catalogs. Red dashed lines in each panel indicate the 1:1 relation, while a black dashed curve shows the best-fit linear relation to the data. We find good agreement between the properties of galaxies in common in both catalogs. The scatter is $\sigma_{M_{\star}} \sim 0.12$, $\sigma_{q} \sim 0.032$ and $\sigma_{R_{\rm{half , \star}}} \sim 0.064$ for the stellar mass, q, and half-mass radius respectively.} 
	\label{fig:app}
\end{figure}

Results in Sec.~\ref{SecSimulations} and, more specifically, in Fig.~\ref{fig:percent}, suggest  that shapes derived for the DESI galaxies are more flattened than those in either GAMA or the ALFALFA survey. A close inspection to Fig.~\ref{fig:distributions_q} reveals that a small bias towards lower $q$ values is already present in the DESI results compared to the other surveys. This difference then accentuates when the $q$ distributions in each survey are used to derive the 3D shapes in Fig.~\ref{fig:percent}.

To inspect this further, we selected a matched sample of galaxies that are present in the DESI and the GAMA catalogs. We select matched galaxies  by their position  in the sky, assuming a tolerance of $3.6$ arcsec between the two catalogs. We show in Fig.~\ref{fig:app} the stellar mass (left), shapes (middle), and half-mass radius, $R_{\rm{half , \star}}$ (right) of these objects as reported for GAMA (x-axis) and DESI (y-axis), finding a tight correlation in both quantities between the two catalogs, with small scatter. The scatter is computed as $\sigma_X = \sqrt{ \frac{1}{N-1} \sum (|X_{\rm DESI} - X_{\rm GAMA}| - \bar{X})^2 }$, where $\bar{X} = \frac{1}{N} \sum |X_{\rm DESI} - X_{\rm GAMA}|$; these corresponds to  $\sigma_{M_{\star}} \sim 0.12$, $\sigma_{q} \sim 0.032$, and $\sigma_{R_{\rm{half , \star}}} \sim 0.064$ for the stellar mass, q, and the half-mass radius respectively. The stellar masses in DESI seem biased high relative to GAMA, but the effect is small. Given our generous stellar mass bins (0.5 to 1 dex, depending on the figure) we do not expect bin swapping to complicate the shape comparison between samples.

Moreover, we find a very good agreement between the projected shape $q$ parameter estimated in DESI and GAMA for each object (right hand panel of Fig.~\ref{fig:app}). This suggests that the larger fraction of flat objects in DESI is unlikely to originate from a systematic bias in the shape, but must instead originate from systematically different populations in DESI vs. GAMA. Since DESI results correspond to the first data release, one may be tempted to consider the GAMA estimates more mature. The additional good agreement between GAMA and ALFALFA in overall shape distributions provides support to the idea of treating estimates from these two surveys as  (conservative) lower limits to the fraction of flat objects, keeping in mind that the fraction could increase if results from DESI are confirmed by future data. 

\section{Simulated galaxy model}
\label{app:app2}

\begin{figure*}
	\centering
	\includegraphics[width=1.0\textwidth]{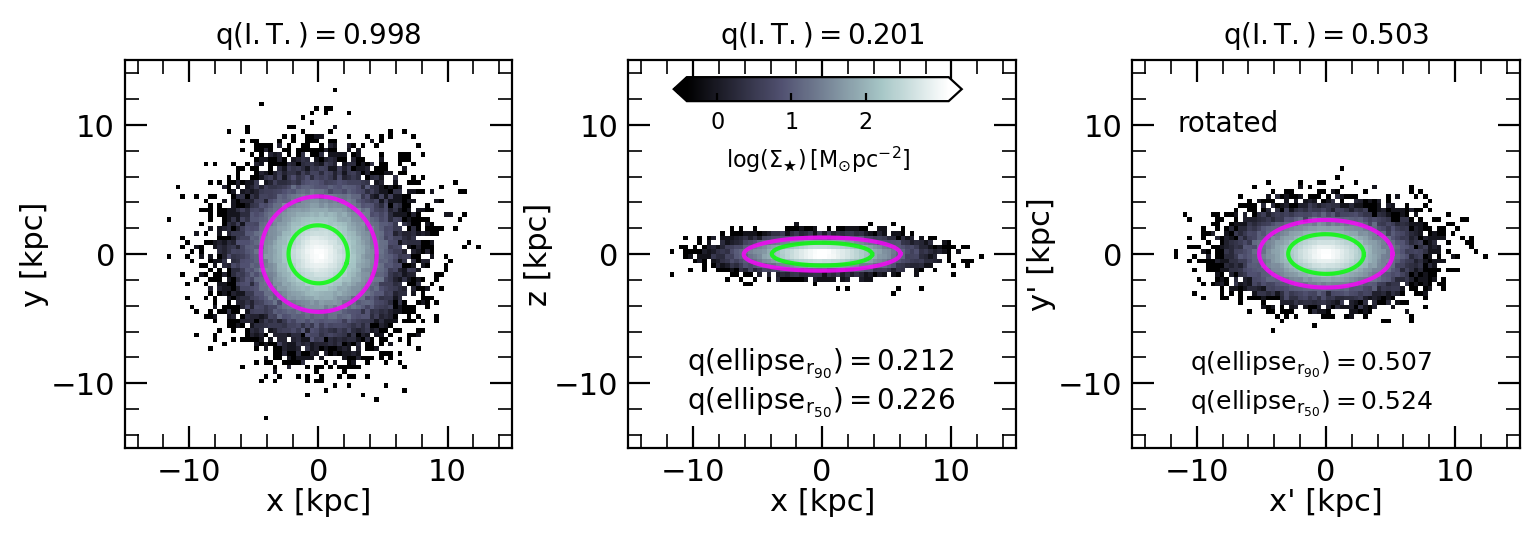}
	\caption{Comparison of q values measured for a simulated galaxy with intrinsic axis ratios $b/a = 1$ and $c/a = 0.2$, generated with $10^5$ stellar particles. The face-on projection (XY) is shown in the left panel, the edge-on projection (XZ) in the middle panel, and an inclined projection in the right panel. At the top of each panel, the axis ratio calculated using the 2D mass-weighted inertia tensor is shown. The green and magenta curves correspond to ellipsoids fitted to isophotes, averaged over all pixels with the same surface mass density, and computed at 50 and 90\% of the total stellar mass ($r_{\rm 50}$ and $r_{\rm 90}$).} 
	\label{fig:app2}
\end{figure*}

To assess how different measurement techniques recover galaxy shapes under various viewing angles, we generate an idealized triaxial galaxy model with a known intrinsic geometry. In Fig.~\ref{fig:app2}, we show a simulated galaxy constructed using 100,000 star particles. In the different panels, we project this model along multiple orientations (face-on, edge-on, and an intermediate inclined view) and measure its apparent shape in each case. Two independent methods are applied to estimate the projected axis ratio ($q$): (i) the 2D mass-weighted inertia tensor, which provides a shape estimate based on the spatial distribution of stellar mass, and (ii) isophotal fitting, where ellipsoids are matched to contours of constant surface mass density chosen to match the radii enclosing 50\% and 90\% of the stellar mass.

While the inertia tensor returns the expected value of the known model, the isophotal method tends to show a strong dependence on the exact radius at which this is measured and to deviate from the known model, particularly in the flatter (i.e., edge-on) projections.

For example, for a model galaxy constructed with intrinsic axis ratios $a=b$ and $c/a=0.2$, as in  Fig.~\ref{fig:app2}, we derived the 2D projected axis ratios from the eigenvalues of the mass-weighted inertia tensor. As expected, this method recovers $q=0.998$ for the face-on view, $q=0.201$ for the edge-on projection, and  $q=0.503$ for the system inclined by 62 degrees (where the expected value is $q=0.5$). In contrast, when we calculate the projected axis ratio from a fitted ellipse, through the isophotal method, the precision depends strongly of the inclination of the system, and the mass surface density ($\Sigma_\star$) used. In the case of $\Sigma_\star$ at the 90\% of the stellar mass ($r_{90}$) we estimated values of $q=0.507$ for the 62 degrees rotated system (i.e., a 1.4\% error) and $q=0.212$ in the edge-on case (6.0\% error). The agreement with the real axis ratios worsens for smaller radii or, equivalently, higher surface brightness cuts. Overall, the inertia tensor method returns closer values to the idealized models, and we use it to compute projected shapes throughout the paper.

\section{Notes on numerical resolution}
\label{app:resolution}

The intrinsic shapes of simulated galaxies are dictated by the galaxy formation model implemented in each model. In addition, numerical effects, such as the use of a finite gravitational softening, may also impose a limit on how ``thin'' simulated galaxies can be. For the simulations included in our analysis, the gravitational softening, $\epsilon$, ranges from 12 pc (FIREbox) to 300 pc and 350 pc at z=0 for TNG50 and Romulus, respectively.

Taking as an example the public data from the TNG50 simulation, we find that the median stellar half-mass radius range from $\sim 0.75$ kpc at $M_{\star} \sim 10^7 ~ \rm{M_{\odot}}$  to $\sim 10$ kpc at $M_{\star} \sim 10^{11} ~ \rm{M_{\odot}}$. Half-mass radii are therefore substantially larger than $\epsilon$ for our entire sample. Assuming (somewhat conservatively) that the minimum thickness resolved is of order the softening length, this means that $c/a<0.2$ and $0.3$ for MW-mass galaxies  correspond to ``thicknesses'' $\sim 2$-$3$ kpc, again much larger then $\epsilon$. We argue that in the high mass end the lack of intrinsically thin objects is a well-measured effect that is not limited by numerical resolution. Note that \citet{Peebles2020} highlighted related issues in the velocity dispersion and orbital distribution in general for the stars of simulated MW-like galaxies using also zoom-in simulations with even higher resolution than TNG50. 

A similar argument can also be made for dwarf galaxies. For $M_{\star}\sim 10^9 ~ \rm{M_{\odot}}$, the average stellar half-mass radii are $\sim 2$ kpc, meaning that $c/a<0.3$ corresponds to thickness $600$ pc, formally sufficiently resolved in the simulation ($\sim 2 \times \epsilon$). At $c/a<0.2$, the half-height needed to be resolved matches the scale of a single softening length in this simulation. For objects as small as $M_{\star} \sim 10^8 ~ \rm{M_{\odot}}$, with stellar half-mass radii $\sim 1$ kpc, the resolution should allow in this simulations to find objects with $c/a<0.3$ rather confidently. Below this scale, however, the frequency of thin dwarfs in TNG50 might be impacted by numerical effects including not only gravitational softening, but also redistribution of mass due to 2-body scattering due to the more massive DM particles \citep{Ludlow2023}. Results presented in our lowest mass bin might require further inspection with future increased resolution simulations. Note that for FIREbox, the substantially smaller softening, 12 pc, means that the vertical structure of all dwarfs within the analyzed range here $M_{\star}=[10^7,10^{11}] ~ \rm{M_{\odot}}$ should be sufficiently resolved.

\section{Accuracy of the deprojection method}
\label{app:app3}

We have tested the accuracy of the deprojection method presented in  Fig.~\ref{fig:percent} using the TNG50 simulation. To this end, we calculate the fraction of thin galaxies using their known 3D c/a measurements (dotted orange lines). We can then compare this to the recovered fraction of galaxies thinner than an intrinsic $c/a$ when starting from the $q$ distributions, as described in Sec. 4, and done for observations and simulations in this paper (long dashed orange lines). We find excellent agreement between both methods, further supporting the reliablity of our results. 

\begin{figure*}
	\centering
	\includegraphics[width=0.45\columnwidth]{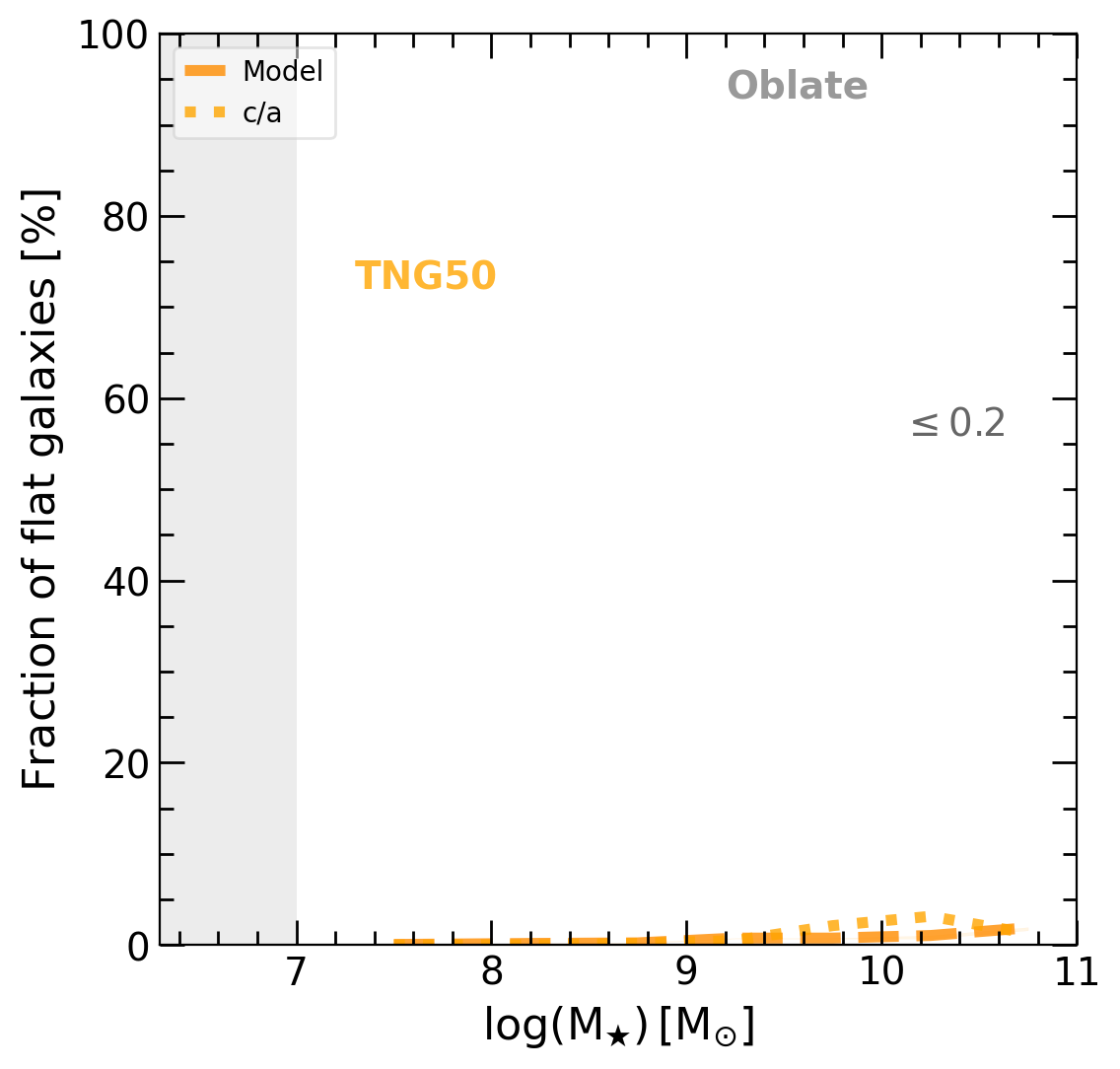}
	\includegraphics[width=0.45\columnwidth]{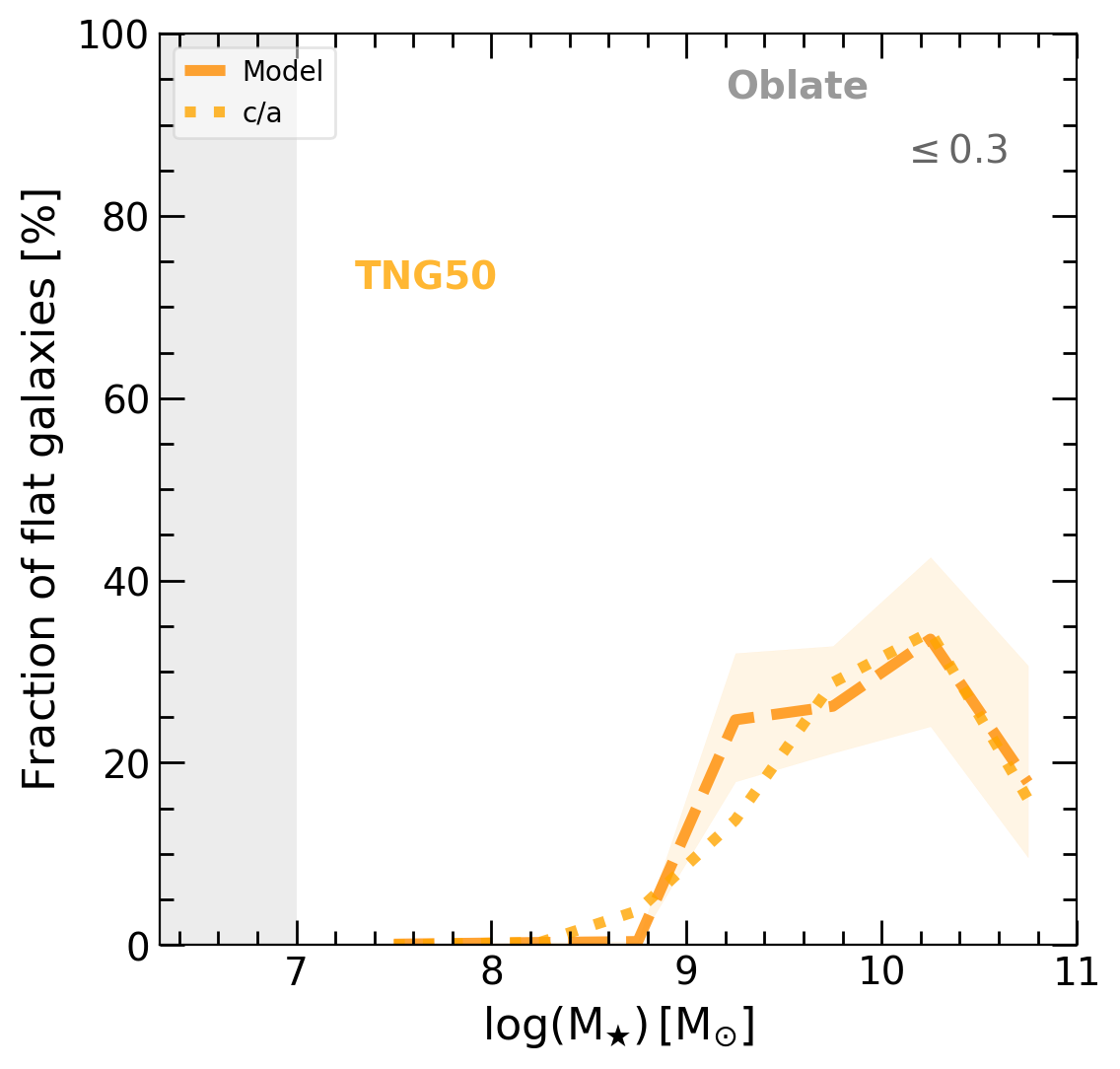}
	\caption{Comparison between the fraction of flat galaxies expected using the deprojection method (dashed orange line) and the known real fraction calculated from the intrinsic c/a of the simulated central galaxies in TNG50 (dotted line).} 
	\label{fig:app3}
\end{figure*}

\label{lastpage}
\end{document}